\documentclass[UTF8]{llncs}

\usepackage{amsmath,amssymb,amsfonts}
\usepackage{graphicx}
\usepackage{hyperref}
\usepackage[braket, qm]{qcircuit}
\usepackage{caption}
\usepackage{subcaption}
\usepackage{color}
\usepackage{mathrsfs}
\usepackage{algorithm}
\usepackage{algorithmic}
\usepackage{booktabs} 
\usepackage{multirow}
\usepackage{array}
\usepackage{makecell}
\usepackage{lineno}
\usepackage{array} 
\usepackage{multirow}
\usepackage{makecell}
\usepackage{graphicx}
\usepackage{adjustbox}
\usepackage{bbding}
\usepackage[misc]{ifsym}
\usepackage{cite}
\usepackage[title]{appendix}
\usepackage{braket}
\usepackage{threeparttable}
\usepackage[section]{placeins}
\begin{document}
\title{Quantum Key Recovery Attack on SIMON Block Cipher}
\author{Hui Liu$^{1,2}$,Li Yang$^{1,2}$\thanks{Corresponding author. E-mail: yangli@iie.ac.cn}}
\institute{$^1$ State Key Laboratory of Information Security, Institute of Information Engineering, Chinese Academy of Sciences, Beijing, China \\
 $^2$ School of Cyber Security, University of Chinese Academy of Sciences, Beijing, China
 }
\maketitle	
	\begin{abstract}
		The quantum security of lightweight block ciphers is receiving more and more attention. However, the existing quantum attacks on lightweight block ciphers mainly focused on the quantum exhaustive search, while the quantum dedicated attacks combined with classical cryptanalysis methods haven't been well studied. In this paper, we study quantum key recovery attack on SIMON block cipher using Quantum Amplitude Amplification algorithm in Q1 model. At first, we reanalyze the quantum circuit complexity of quantum master key exhaustive search on SIMON block cipher. The Clifford gates count is estimated more accurately and the T gate count is reduced. We also reduce the T-depth and Full-depth due to some minor modifications to the circuit. Then, based on the differential cryptanalysis on SIMON32, SIMON48 and SIMON64 given by Biryukov et al. in FSE 2014, we give quantum round key recovery attacks on these SIMON variants and analyze quantum circuit complexity separately. We take the quantum attack on 19-round SIMON32/64 for an example and design the quantum circuit of the key recovery process. The two phases of this attack could be regarded as two QAA instances separately, and the first QAA instance consists of four sub-QAA instances. We conclude that the encryption complexity and circuit complexity of quantum dedicated attacks on 19-round SIMON32/64, 19-round SIMON 48 and 26-round SIMON64/128 are both lower than those of the quantum exhaustive search on these variants separately. Our work firstly studies the quantum dedicated attack on SIMON block cipher from the perspective of quantum circuit complexity, which is a more fine-grained analysis of quantum dedicated attacks' complexity.
	\end{abstract}
	
	\noindent{\bf Keywords:} Quantum cryptanalysis; Quantum Amplitude Amplification algorithm; Differential cryptanalysis; Key recovery attack; SIMON block cipher

	\section{Introduction}
\label{sect:intro}
The development of quantum computation poses a threat to classical cryptosystems. Shor's algorithm \cite{shor94} can break the security of public-key cryptosystems based on integer factorization and discrete logarithm, which gives rise to post-quantum cryptography. As for the symmetric cryptosystems, before Simon's algorithm \cite{simon97} is applied in quantum cryptanalysis, there is only Grover's algorithm \cite{gro97} that helps get a quadratic speed-up.\\ 
\indent Quantum cryptanalysis against block ciphers receives much attention in recent years. Following the notions for PRF security in quantum setting proposed by Zhandry et al. \cite{zha12}, there are two security models in quantum cryptanalysis against block ciphers, called Q1 model and Q2 model by Kaplan et al. in \cite{kap16a}.\\
\textbf{Q1 model:} The adversary is only allowed to make classical online queries and do quantum offline computation.\\
\textbf{Q2 model:} The adversary is allowed to do offline quantum computation and make online quantum superposition queries. That is, the adversary could query in a superposition state to the oracle and get a superposition state as a query result.\\
\indent We can observe that Q1 model is more realistic than Q2 model for the reason that it's up to the oracle whether to allow superposition access. However, it's still meaningful to study Q2 model to prepare for the future with highly developed quantum communication technology.\\
\indent Quantum cryptanalysis in Q2 model has been going on for a long time. In 2010, Kuwakado and Morii constructed a quantum distinguisher on 3-round Feistel structure \cite{km10} using Simon's algorithm in Q2 model. Then they recovered the key of Even-Mansour also using Simon's algorithm in \cite{km12}. At Crypto2016, Kaplan et al. extended the result in \cite{km10,km12} and applied Simon's algorithm to attack a series of encryption modes and authenticated encryption such as CBC-MAC, PMAC, OCB \cite{kap16b}. In Q2 model, Simon's algorithm can be combined with Grover's algorithm to attack block ciphers. Leader and May \cite{lm17} firstly used this idea to attack FX construction in Q2 model. Inspired by this work, Dong et al. \cite{dong20} gave a quantum key recovery attack on full-round GOST also in Q2 model. Besides, Bernstein-Vazarani (BV) algorithm \cite{bv97} can also be applied in quantum cryptanalysis. Li and Yang \cite{ly15} proposed two methods to execute quantum differential cryptanalysis based on BV algorithm. Then, Xie and Yang extended the result in \cite{ly15} and present several new methods to attack block ciphers using BV algorithm \cite{xy19}.\\
\indent In Q1 model, it seems as if quantum cryptanalysis becomes less powerful. The most trivial quantum attack is the quantum exhaustive search that defines the general security of block ciphers in the quantum setting. Grassl et al. present quantum circuits to implement an exhaustive key search on AES and estimate quantum resources  in Q1 model \cite{glr16}. After that, there are also some other results exploring the quantum circuit design of AES \cite{asa18,jnr20,zou2020quantum,ls20}. Besides, there are many attempts of quantum dedicated attacks combined with classical cryptanalysis methods, e.g. differential and linear cryptanalysis \cite{kap16a}, meet-in-the-middle attacks \cite{hs18,bns19}, and rebound attacks \cite{hosoyamada2020finding,dong2020quantum}. In addition, Grover's algorithm is not the only algorithm used for quantum cryptanalysis in Q1 model. Based on the work of Leader and May in \cite{lm17}, Xavier et al. used the special algebraic structure of some block ciphers to give the attack using offline Simon's algorithm in Q1 model for the first time\cite{bonnetain2019quantum2}. Then, Xavier Bonnetain et al. present the first complete implementation of the offline Simon’s algorithm and estimate its cost to attack in \cite{bonnetain2020quantum}, which measures the quantum dedicated attack's complexity from circuit complexity.\\ 
\indent The research of lightweight block ciphers has received much attention in a decade. Several lightweight primitives have been proposed by the researchers, to just name some, SIMON\cite{bss15}, SPECK\cite{bss15}, SKINNY\cite{bjk16}, PRESENT\cite{bkl07}. To prepare for the future with large-scale quantum computers, it's necessary to study the quantum security of lightweight block ciphers. There are several attempts to study the quantum generic attacks on some lightweight block ciphers \cite{amm20,jck20,rmm20}. In this paper, we focus on the quantum security of SIMON. The family of SIMON algorithm \cite{bss15} is a lightweight block cipher proposed by NSA in 2013, which has outstanding hardware implementation performance. In the classical setting, there have been many dedicated attacks aimed at SIMON. However, in the quantum setting, the only quantum attack on SIMON is in \cite{amm20} where Anand et al. present the quantum circuit of Grover's algorithm on SIMON variants and give corresponding quantum resources estimate, which is a quantum generic attack. To further explore the quantum security of SIMON, we need to study the quantum dedicated attacks on SIMON. Notably, when measuring the attack complexity, the existing quantum dedicated attacks all only focused on encryption complexity except the result in \cite{bonnetain2020quantum}, while we use the quantum circuit complexity to measure the complexity of quantum dedicated attacks in our study.\\

\noindent\textbf{Attack Model} We consider the chosen-plaintext attack on SIMON block cipher in Q1 model, where the adversary is allowed to make classical online queries of encryption oracle and can choose random message pairs with input differential $\Delta{x}$. To achieve such an attack, the adversary needs to implement transformation:
$$\sum_{i=1}^{q}\ket{0}\ket{0}\ket{0}\rightarrow\sum_{i=1}^{q}\ket{m_i}\ket{0}
\ket{0}\rightarrow\sum_{i=1}^{q}\ket{m_i}\ket{E(m_i)}
\ket{E(m_i\oplus \Delta{x})}$$ 
when given $q$ pairs of classical plaintext-ciphertext pair as input. We suppose this process is efficient. Thus we can ignore the quantum circuit complexity of this process.\\

\noindent\textbf{Our Contribution} In this paper, we study the quantum key recovery attack on SIMON block cipher using Quantum Amplitude Amplification algorithm(QAA) in Q1 model. Our contributions can be summarized in the following two aspects.
\begin{enumerate}
	\item We reanalyze the quantum circuit complexity of quantum master key search on SIMON block cipher. On one hand, we give a more accurate estimate result of Clifford gates count and reduce T gate count. We reduce the execution number of the key expansion process, which brings down the number of NOT gates and CNOT gates. Besides, counting the Clifford gates decomposed by Toffoli gates and multi controlled-NOT into the total number of Clifford gates helped us give a more accurate estimate of Clifford gates count. And we reduce the number of T gates using the decomposition of multi-control NOT gates with auxiliary qubits. On the other hand, we give a more thorough analysis of circuits' depth. The depth we focus on here is the depth of such quantum circuits that only are composed of Clifford gates and T gates. We make some modifications to the circuit of SIMON block cipher's round function, which reduces the Full-depth. Compared to \cite{amm20}, the circuit complexity of quantum master key search in our estimate is more accurate and thorough. 
	\item We combine the classical differential cryptanalysis to give the quantum key recovery attack on 19-round SIMON32/64, 19-round SIMON48 and 26-round SIMON64 for the first time. We take the quantum round key recovery attack on 19-round SIMON32/64 for an example and design the quantum circuit of the attack process, whose two stages, the quantum partial key guessing stage and the quantum remaining keys search stage, can be regarded as two QAA instances. The first instance includes four sub-QAA instances, which respectively correspond to the four processes of using four differentials for key recovery. We found that except for the quantum dedicated attack on 26-round SIMON64/96, the encryption complexity and quantum circuit complexity of the quantum dedicated attacks on other SIMON variants given in this article are both lower than those of the quantum master key exhaustive search attack. The previous work almost only studied the encryption complexity of quantum dedicated attacks. However, our work studied the quantum circuit complexity of quantum dedicated attacks, which is a more fine-grained perspective.
\end{enumerate}

\noindent\textbf{Outline} The rest of the paper is organized as follows. In Section \ref{sec:pre}, we introduce the notations used in this paper and the background knowledge of SIMON block cipher, QAA algorithm and quantum circuit. In Section \ref{sec:qmks}, we reanalyze the quantum circuit complexity of quantum master key exhaustive search on SIMON32/64. In Section \ref{sec:qrkr}, we describe the quantum round key recovery attack on 19-round SIMON32/64. In Section \ref{sec:complexity}, we compare the complexity of quantum master key search attack and quantum round key recovery attack. In Section \ref{sec:conclusion}, we make a summary of this paper. In Appendix \ref{sec:app}, we list the circuit complexity of the quantum attacks on SIMON48 and SIMON64.

	\section{Preliminaries}
\label{sec:pre}
\subsection{Notations}
For convenience, we list the notations used in this paper in Table \ref{tb:notations}.
\begin{table}[!htbp]
	\caption{\textmd{Notations}}
	\renewcommand\arraystretch{1.3}
	\resizebox{\textwidth}{!}{
		\begin{tabular}{ll}
			\hline
			Notation & Description\\\hline
			$\&$ & The bitwise AND operation\\
			$\oplus$  & The bitwise XOR operation\\
			$\lll$ & The cyclic left rotation operation\\
			Round-$i$ & The $i$-th round of SIMON block cipher\\
			$(L^{i-1},R^{i-1})$ & The input block of Round-$i$ in SIMON block cipher\\
			$L^i[j]$ & The $j$-th bit of $L^i$(the index of rightmost bit is 0)\\
			$K^{i-1}$ & The round key of Round-$i$ in SIMON block cipher\\
			$\Delta^{i-1}=(\Delta L^{i-1},\Delta R^{i-1})$ & The input difference of Round-$i$ in SIMON block cipher\\
			$\Delta And^{i}$ & $\Delta And^{i}:=(L^i \lll 1) \& (L^i\lll8)\oplus((L^{i})^{\prime} \lll 1) \& ((L^{i})^{\prime} \lll 8)$\\
			$\Delta Rot^{i}$ &  $\Delta Rot^{i}:=\Delta L^i\lll2$\\
			$E(\cdot)$ & The encryption function of 19-round SIMON32/64 with real key $k$\\
			$E_k(\cdot)$ & The encryption function of 19-round SIMON32/64 with guessed key $k$\\
			$D_k^{j}(\cdot)$ & The decryption function that decrypts the given ciphertext in $j$ rounds with key $k$\\
			$\mathcal{QMKS}$ & The quantum master key exhaustive search attack on SIMON block cipher\\
			$\mathcal{QRKR}$ & The quantum round key recovery attack on SIMON block cipher\\
			\#iter & The number of iteration in a QAA instance\\
			\#Toff-C & The number of CNOT gate decomposed by Toffoli gate\\
			\#Toff-H & The number of H gate decomposed by Toffoli gate\\ 
			\#Toff-S & The number of S gate decomposed by Toffoli gate\\
			
			\hline
			\label{tb:notations}
		\end{tabular}
	}
\end{table}

\subsection{Brief Description of SIMON}
SIMON is a Feistel structure lightweight block cipher. There are many SIMON variants to adapt to different computing scenarios, the differences between which lie at block size, key size, word size and round number. The block size of SIMON is $2n$ bits while the key size is $mn$ bits. We could use SIMON$2n/mn$ to denote all SIMON variants, where $n\in\{16,24,32,48,64\}$ and $m\in\{2,3,4\}$. In addition, we can also use SIMON$2n$ to denote all the SIMON variants with $2n$-bit block size. All the SIMON variants are summarized in Table \ref{tb:simonvar}.\\
\begin{table}[!htbp]
	\caption{\textmd{All SIMON variants}}
	\renewcommand\arraystretch{1.3}
	\resizebox{\textwidth}{!}{
		\begin{tabular}{ccccc}
			\hline  
			Block Size(2$n$)&Key Size($k=mn$)&Word Size($n$)&Key Words($m$)&Rounds($T$) \\\hline
			
			32&64&16&4&32 \\
			
			48&72,96&24&3,4&36,36 \\
			
			64&96,128&32&3,4&42,44 \\
			
			128&128,192,256&64&2,3,4&68,69,72 \\
			\hline  
			\label{tb:simonvar}
		\end{tabular}
	}
\end{table}

\begin{figure}[!htbp]
	\centering
	\includegraphics[scale=0.7]{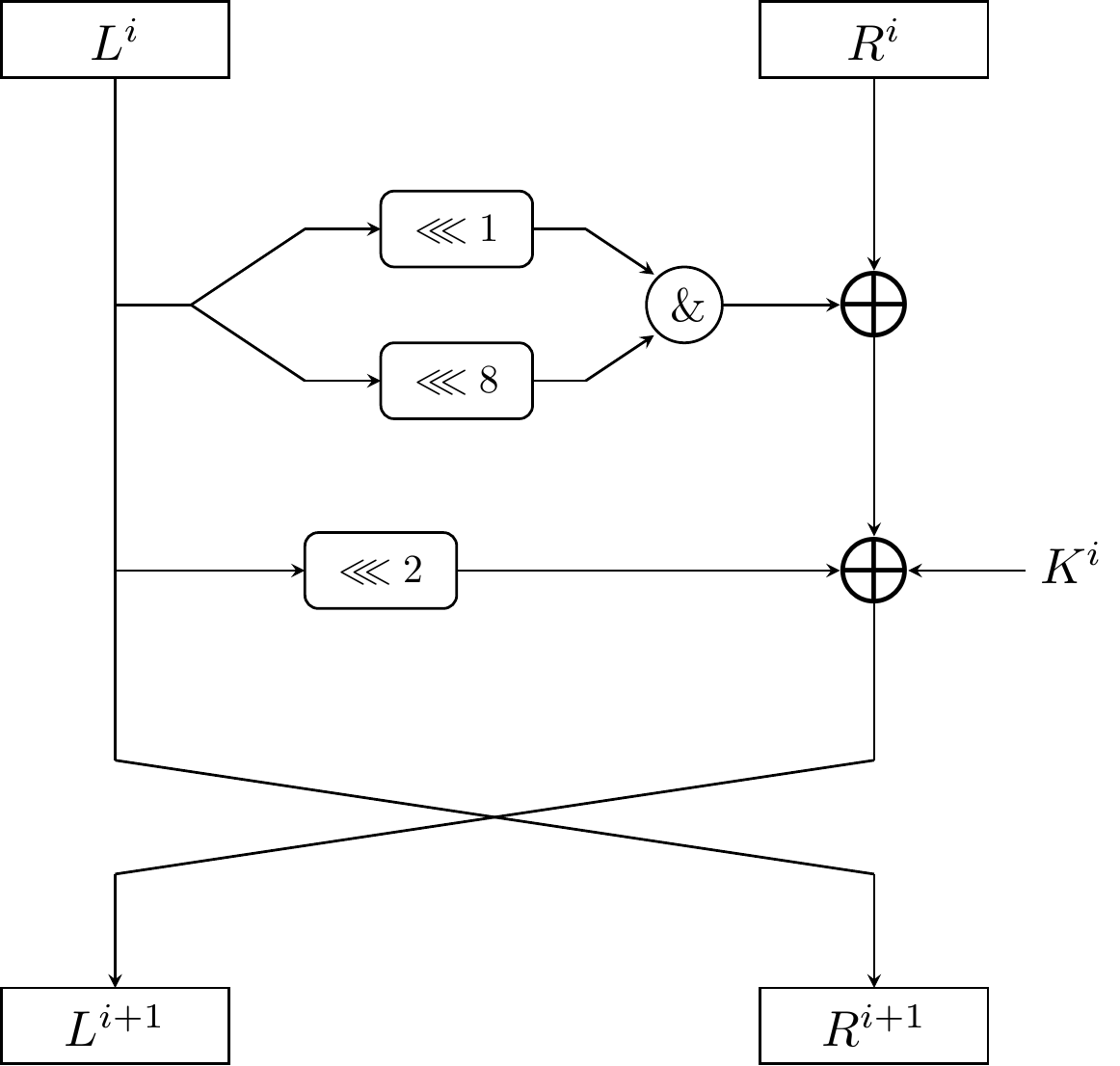}
	\caption{Round function of SIMON}
	\label{fig:roundfunc}
\end{figure}

\noindent \textbf{Round Function} The $i$-th iteration structure of SIMON$2n/mn$ is shown in Fig. \ref{fig:roundfunc}. We can easily see that the round function of SIMON$2n/mn$ only consists of bit-wise AND, cyclic left rotation and bit-wise XOR. For SIMON block cipher, $f:\{0,1\}^n\rightarrow\{0,1\}^n$ is defined as $f(x) = (x \lll1)\&(x\lll8) \oplus (x\lll2) $. The round function of SIMON block cipher is defined as follows:
\begin{equation*}
	\begin{aligned}
		F:\{0,1\}^{n} \times\{0,1\}^{n} &\rightarrow \{0,1\}^{2n}\\
		F(L^{i},R^{i}) &= (R^{i}\oplus f(L^{i})\oplus K^{i},L^{i})	
	\end{aligned}
\end{equation*}

\noindent \textbf{Key Schedule} For $r$-round SIMON$2n/mn$, the round key of SIMON is derived from primary key $\{K^0,K^1,\cdots,K^{m-1}\}$. When $i = 0,1,\cdots,m-1$, $K^i = K^i$. When $i = m,m+1,\cdots,r-1$, the specific key expansion scheme is defined as follows.
\begin{enumerate}
	\item If $m=2$, $K^i = c \oplus (z_j)^{i-m}\oplus K^{(i-m)} \oplus (K^{(i-m+1)}\ggg3)\oplus(K^{(i-m+1)}\ggg4)$;
	\item If $m=3$, $K^i = c \oplus (z_j)^{i-m}\oplus K^{(i-m)} \oplus (K^{(i-m+2)}\ggg3)\oplus(K^{(i-m+2)}\ggg4)$;
	\item If $m=4$, $K^i = c \oplus (z_j)^{i-m}\oplus K^{(i-m)} \oplus K^{(i-m+1)}\oplus(K^{(i-m+1)}\ggg1)\oplus(K^{(i-m+3)}\ggg3)\oplus(K^{(i-m+3)}\ggg4)$;
\end{enumerate}
$z_j$ is a constant sequence and $c = 2^n - 4$. The key schedule is linear. Thus we can derive the master key from any $mn$ independent bits of subkeys. Particularly, for SIMON32/64, as long as we get the round keys of any four adjacent rounds, the master key can be easily deduced.\\

\noindent \textbf{Related Works} In the classical setting, there already have been some attack results on SIMON block cipher. We make a simple summary of some attacks on SIMON block cipher in Table \ref{tb:attacks}. However, in the quantum setting, the only quantum attack on SIMON is the quantum exhaustive search in \cite{amm20}. To further explore the quantum security of SIMON block cipher, it's necessary to study the quantum dedicated attack on SIMON, which is our focus in this paper. \\
\begin{table}[!htbp]
	\caption{Summary of some classical attacks on SIMON block cipher}
	\footnotesize
	\renewcommand\arraystretch{1.5}
	\resizebox{\textwidth}{!}{
		\begin{tabular}{ccccccc}
			\hline
			Cipher & Round & Attacked Round & Technique & Time & Data &  Reference  \\
			\hline
			\multirow{3}{*}{{SIMON32/64}}

			& 32 &  $19$ & Differential & $2^{34}$ & $2^{31}$  & \cite{brv14} \\
			& 32 &  $21$ & Differential & $2^{55.25}$ & $2^{31}$  & \cite{wn18} \\
			& 32 &  $21$ & Linear & $2^{60.99}$ & $2^{28.99}$  & \cite{shs17} \\
			\hline
			\multirow{3}{*}{{SIMON48/72}}
			& 36 &  $19$ & Differential & $2^{52}$ & $2^{46}$ & \cite{brv14} \\
			& 36 &  $23$ & Linear & $2^{65.34}$ & $2^{47.92}$ & \cite{cw16} \\
			& 36 &  $24$ & Integral & $2^{71}$ & $2^{48}$ & \cite{chu2018improved} \\
			\hline
			
			\multirow{3}{*}{{SIMON48/96}}
			& 36 &  $19$ & Differential & $2^{69}$ & $2^{46}$ & \cite{brv14} \\
			& 36 &  $22$ & Zero-correlation & $2^{80.5}$ & $2^{48}$ & \cite{wlz14}\\
			& 36 &  $25$ & Linear & $2^{88.28}$ & $2^{47.92}$ & \cite{cw16} \\
			\hline	
			
			\multirow{3}{*}{{SIMON64/96}}	
			& 42 &  $26$ & Differential & $2^{90.4}$ & $2^{64}$ & \cite{brv14} \\
			& 42 &  $30$ & Linear & $2^{88.13}$ & $2^{63.53}$ & \cite{cw16} \\
			& 42 &  $30$ & Differential & $2^{88}$ & $2^{63.3}$ & \cite{wn18} \\
			
			\hline
			
			\multirow{3}{*}{{SIMON64/128}}
			& 44 &  $26$ & Differential  & $2^{121}$ & $2^{63}$ & \cite{brv14} \\
			& 42 &  $31$ & Linear & $2^{120}$ & $2^{63.53}$ & \cite{cw16} \\
			& 44 &  $31$ & Differential  & $2^{120}$ & $2^{63.3}$ & \cite{wn18} \\
			\hline
			
			\label{tb:attacks}
		\end{tabular}
	}
\end{table}

\subsection{Brief Description of QAA algorithm}
\begin{figure}[!htbp]
	\centering
	\includegraphics[width=1\textwidth]{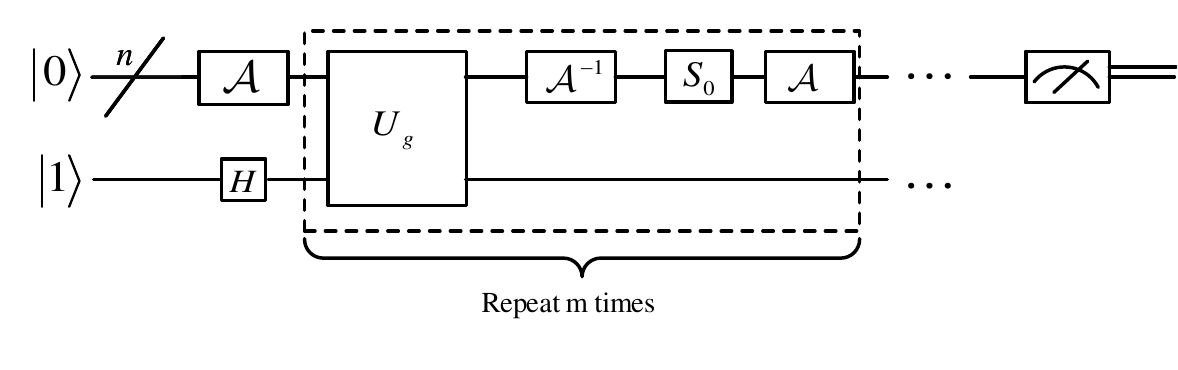}
	\caption{Quantum circuit of QAA algorithm}
	\label{fig:qaa}
\end{figure}

Grover proposed a quantum search algorithm in \cite{gro97} used for finding some specific element in the database containing $N$ elements. Compared to classical search algorithms, Grover's algorithm could bring quadratic speed-up and its optimality was proved in \cite{zalka1999grover}. Then Brassard et al. further extended Grover's algorithm and proposed Quantum Amplitude Amplification algorithm(QAA). QAA algorithm can be seen as a generalization of Grover's algorithm which could also achieve quadratic speed-up compared to classical algorithms. The quantum circuit of QAA algorithm is shown in Fig. \ref{fig:qaa}. According to \cite{bhm02}, QAA algorithm can be summarized as Theorem \ref{th:th1}.
\begin{theorem}\label{th:th1}
	Let $\mathcal{A}$ be any quantum algorithm that uses no measurements, and $g:\{0,1\}^n\rightarrow\{0,1\}$ be any Boolean function. The positive number $p$ is the initial success probability of $\mathcal{A}$. We define $G=U_sU_g=-\mathcal{A}S_0\mathcal{A}^{-1}U_g$, where $S_0= 2\ket{0}\bra{0}-I$. If we compute $G^m \mathcal{A}\ket{0}$ where $m=\lfloor \frac{\pi}{4\arcsin{\sqrt{p}}} \rfloor$ and then measure the system, the outcome is good with probability at least $max(1-p,p)$.
\end{theorem}

\indent For simplicity, we call a search problem using QAA algorithm to settle as a QAA instance. The iteration operator of a QAA instance is then called QAA iterator. For a QAA instance with $M$ solutions in $N=2^n$ elements, we define elements that are solutions as \emph{GOOD} while the elements that are not solutions as \emph{BAD}.  We define a function $g:\{0,1\}^n\rightarrow\{0,1\}$ 
\begin{eqnarray*}
	\centering
	g(x)=
	\begin{cases}
		1,\text{if} \;  x \; \text{is \emph{GOOD}}\\
		0,\text{if} \;  x \; \text{is \emph{BAD}}\\
	\end{cases}
\end{eqnarray*}

\indent According to the computation result of function $g$, we can divide $N$ elements into two collections, $A=\{x\in\{0,1\}^n,\text{if}\quad g(x)=1\}$ and $B=\{x\in\{0,1\}^n,\text{if} \quad g(x)=0\}$. The superposition state of $N$ elements $\ket{s}=\sum_{x=0}^{2^n-1}\alpha\ket{x}$ could be written as $\ket{s}=\ket{\psi_0}+\ket{\psi_1}$, where $\ket{\psi_1}$ represents the superposition state of all computational basis states that make the output of $g$ is 1 and $\ket{\psi_0}$ is the superposition state of all computational basis states that make the output of $g$ is 0. Based on function $g$, we construct operator $U_g$ only used for change the amplitude sign of all \emph{GOOD} elements, which is defined as
\begin{eqnarray*}
	U_g\ket{x}=
	\begin{cases}
		-\ket{x},\text{if} \;  x \; \text{is \emph{GOOD}}\\
		\ket{x},\text{if} \;  x \; \text{is \emph{BAD}}\\
	\end{cases}
\end{eqnarray*}

\indent QAA algorithm amplifies the amplitude of \emph{GOOD} elements through the following process.
\begin{enumerate}
	\item Apply $\mathcal{A}$ on the initial state $\ket{\psi}=\ket{0}$, we can get  	$$\ket{\psi}=\mathcal{A}\ket{0}=\ket{\psi_0}+\ket{\psi_1}=\sum_{x\in A}\alpha_x\ket{x}+\sum_{y\in B}\alpha_y\ket{y}$$
	$p=\braket{\psi_1|\psi_1}$ is the probability of getting a GOOD element as well as the initial success probability.
	\item Call QAA iteration $m=\lfloor \frac{\pi}{4\arcsin{\sqrt{p}}} \rfloor$ times. In each iteration, there are two steps.
	The first step is to apply $U_g$ to the quantum state $\ket{\psi}$, after which we can get $$U_g\ket{\psi}=\ket{\psi_0}-\ket{\psi_1}$$ The second step is to apply diffusion operator $2\ket{s}\bra{s}-I$ to $\ket{\psi}$, where $\ket{s}$ is the superposition of all $N$ elements.
	\item Measure the first register and we can obtain one of the \emph{GOOD} elements with probability close to 1.	 	
\end{enumerate}

\indent We can observe that compared to the original Grover's algorithm, QAA algorithm not only ensures the quadratic speed-up but also is more universal. In Grover's algorithm, the operator $\mathcal{A}$ used for creating initial state can only be $H^{\otimes n}$. That is, we can only search from the equal superposition state $\ket{s}=\frac{1}{\sqrt{N}}\sum_{x=0}^{N-1}\ket{x}$. As a result, QAA algorithm is more suitable for our attack scenario. It is worth noting that we must carry out plenty of measurements to get all solutions because the output of QAA algorithm is the superposition of $M$ solutions.\\
\subsection{Quantum Circuit}
\begin{figure}[!htbp]
	\centering
	\includegraphics[width=1\textwidth]{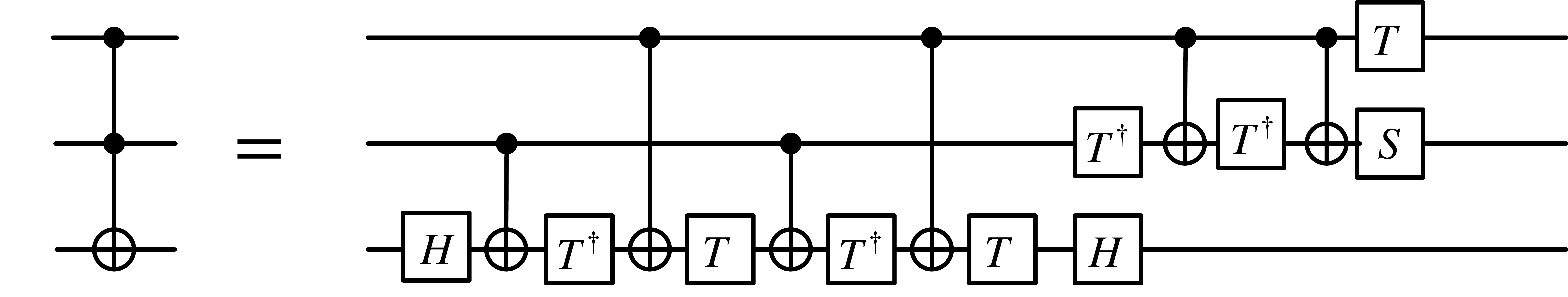}
	\caption{The decomposition of Toffoli gate in \cite{nil}}
	\label{fig:toffoli}
\end{figure}

\begin{figure}[!htbp]
	\centering
	\includegraphics[width=1\textwidth]{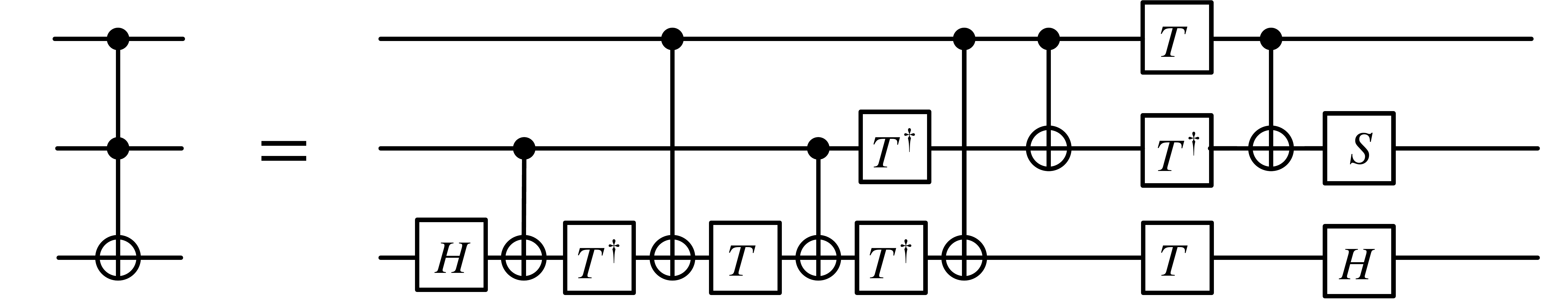}
	\caption{The decomposition of Toffoli gate in \cite{mdm13}}
	\label{fig:toffoli2}
\end{figure}

Quantum logic gates are the foundation of quantum circuits. A quantum circuit can be seen as a sequence of quantum logic gates. In order to measure the complexity of a quantum circuit, we should consider the number of gates, and the number of qubits and the depth. When computing the depth of a quantum circuit, we also adopt the \emph{full parallelism assumption} as in \cite{jnr20}, which means a quantum circuit can apply any number of gates simultaneously so long as these gates act on disjoint sets of qubits.\\
\indent The Clifford + T gate set forms a set of universal quantum gates. The Clifford group is defined as the group of unitary operators that map the group of Pauli operators to itself under conjugation. The Clifford gates are then defined as elements in the Clifford group. The basic Clifford gates set is $\{H,S,CNOT\}$. However, we cannot achieve universal quantum computation only with Clifford gates. This is, non-Clifford gates should be added to the gate set. And T gate is usually the choice to be added in. The matrix representations of Clifford + T gate set in shown in Eq.(\ref{eq:qgate}).\\
\begin{equation}
	\label{eq:qgate}
	\resizebox{.9\hsize}{!}{
		$H=\frac{1}{\sqrt{2}}
		\left(
		\begin{array}{cc}
			1 & 1\\
			1 & 1
		\end{array}
		\right),
		S=
		\left(
		\begin{array}{cc}
			1 & 0\\
			0 & i
		\end{array}
		\right),
		CNOT=
		\left(
		\begin{array}{cccc}
			1 & 0 & 0 & 0\\
			0 & 1 & 0 & 0\\
			0 & 0 & 0 & 1\\
			0 & 0 & 1 & 0
		\end{array}
		\right),
		T=
		\left(
		\begin{array}{cc}
			1 & 0\\
			0 & e^{i\frac{\pi}{4}}
		\end{array}
		\right)	$
	}
\end{equation}
\indent According to \cite{mdm13}, all Clifford group operations have transversal implementations and thus are relatively simple to implement while non-Clifford gates require much more sophisticated and costly techniques to implement. The surface codes, which promise higher thresholds than concatenated code schemes, also have a significantly more complicated T gate implementation than any of the Clifford group generators. As a result, it's significant to study the number of T gate in a quantum circuit in order to measure the complexity of quantum computation. Besides, Amy et al. proposed T-depth as a cost function of quantum circuits in \cite{mdm13}. We can observe that the research on reducing the T-depth of quantum circuits has been paid more and more attention.\\
\indent In classical computation, the Toffoli gate is a universal classical reversible logic gate, while for quantum computation it needs to be decomposed into Clifford gates and T gates for real implementation. According to \cite{nil}, the decomposition of the Toffoli gate is shown in Figure \ref{fig:toffoli} where a Toffoli gate can be decomposed into 7 T gates, 6 CNOT gates, 2 H gates and 1 S gate with T-depth 7 and Full depth 13. Then, to reduce T-depth, Amy et al. proposed a decomposition scheme of Toffoli gate in \cite{mdm13} with T-depth 3 and Full depth 10, shown in Fig. \ref{fig:toffoli2}. And Amy et al. conjectured that this T-depth is optimal for circuits without ancillary qubits. Although T-depth could be reduced to 1 further with ancilla qubits according to Fig. 1 in \cite{selinger2013quantum}, the number of CNOT gates increases much. After an overall consideration of gate counts and T-depth of quantum circuits, we adopt the method in Fig. \ref{fig:toffoli2} to decompose the Toffoli gate in this paper.

\begin{figure}[!htbp]
	\centering
	\includegraphics[width=1\textwidth]{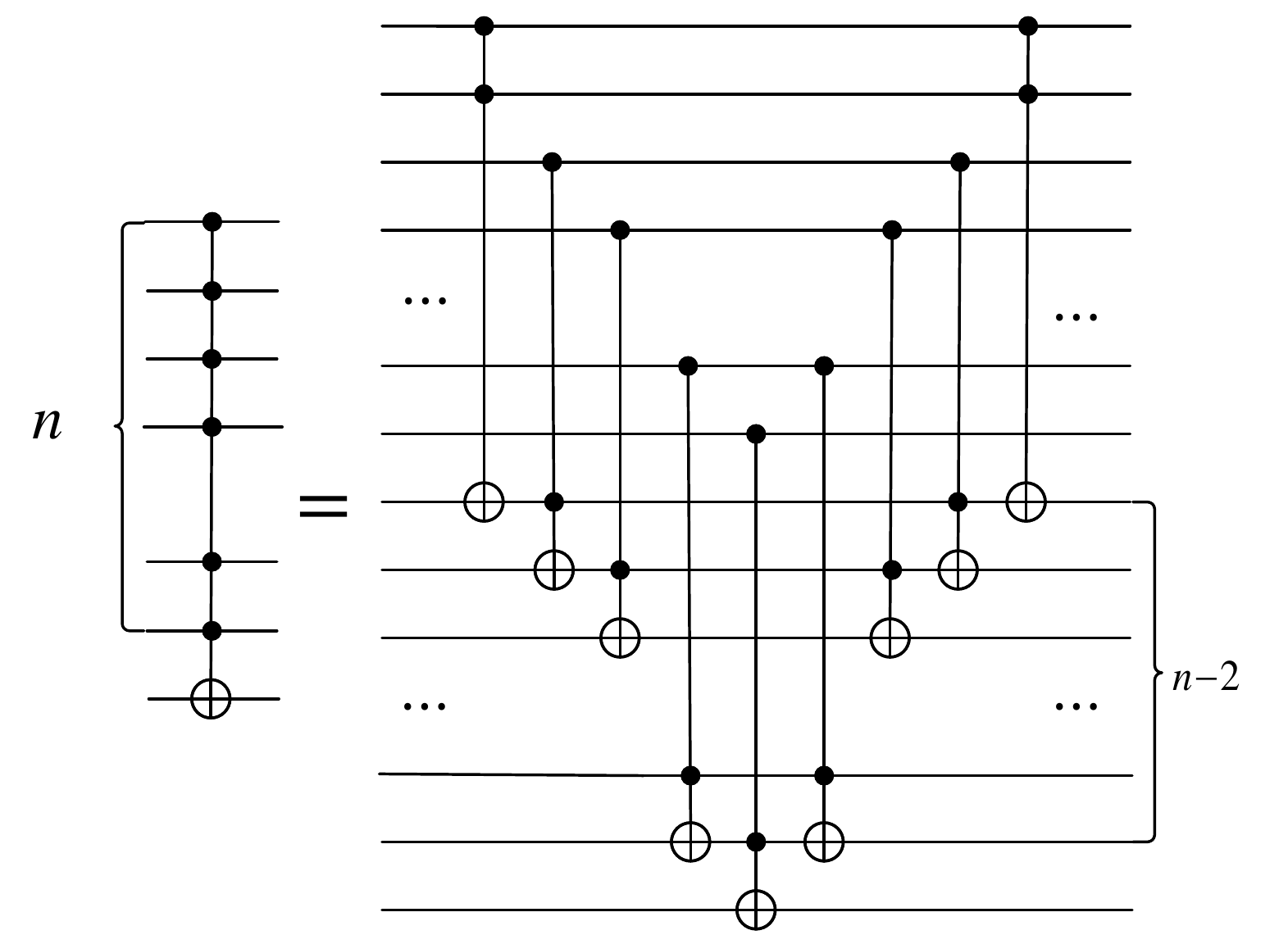}
	\caption{The decomposition of $n$-fold controlled-NOT}
	\label{fig:nfoldCNOT}
\end{figure}

\indent In QAA iterator $G$, there two multi controlled-NOT gates. For the real implementation of QAA algorithm, we need to decompose the multi controlled-NOT gate into a series of Toffoli gates. Then we need to decompose the Toffoli gate into Clifford + T gates. According to \cite{nil}, the $n$-fold controlled-NOT could be decomposed into $2n-3$ Toffoli gates using $n-2$ ancilla qubits. We show the decomposition of $n$-fold controlled-NOT in Fig. \ref{fig:nfoldCNOT}. Here, we offer a concept, Toffoli-depth, which is similar to T-depth, meaning the number of stages in the circuit involving Toffoli gates. In our analysis, computing the Toffoli-depth is the first step to compute the T-depth and Full depth of quantum circuits. We can observe that the Toffoli-depth of Fig. \ref{fig:nfoldCNOT} is $2n-3$. Thus the full depth of implementing a $n$-fold controlled-NOT is $20n-30$ ,and the T-depth is $6n-9$. It is worth noting that the depth we're talking about refers to the depth of the quantum circuits only containing Clifford gates and T gates. This is, we need to decompose all Toffoli gates into Clifford gates and T gates before computing the depth of quantum circuits.

	\section{The Quantum Master Key Exhaustive Search Attack on 19-round SIMON32/64}
\label{sec:qmks}
In this section, we reanalyze the quantum circuit complexity of $\mathcal{QMKS}$ on 19-round SIMON32/64 using QAA algorithm based on the result in \cite{amm20} where Anand et al. present Grover's algorithm on SIMON variants and estimate the quantum circuit complexity to implement such an attack. \\
\subsection{The Quantum Circuit Complexity of SIMON32/64 }\label{sec:simon}
\indent At first, we reanalyze the quantum circuit complexity of implementing 19-round SIMON32/64. From Table 3 in \cite{amm20}, we can easily derive the gate count of implementing 19-round SIMON32/64. However, when computing the circuit depth, we got different results from \cite{amm20}. Anand et al. implemented all SIMON variants in Qiskit\cite{koch2019introduction}. The circuit depth can be calculated using the Qiskit function. After running the code of implementing SIMON32/64 in \cite{quantsimon}, we found that the Qiskit function computes the depth of the quantum circuit without decomposing the Toffoli gate which leads to the incompleteness of the circuit depth calculation. In our estimate, Toffoli gates should be decomposed into Clifford gates and T gates before computing the circuit depth. Besides, we made some small modifications to the code of implementing SIMON32/64, which brought in a reduction of Full-depth. We performed one operation on all bits firstly and then performed the next operation on all bits, instead of performing all operations on each bit one by one in our modifications. We gave our modified code in \cite{simon3264Q}. We list the quantum circuit complexity of implementing SIMON32/64 in Table. \ref{tb:simonimpres}. In addition, we did a similar analysis on the quantum circuits of SIMON48 and SIMON96 and listed the circuit complexity in Table \ref{tb:qmkrressum3} in Appendix \ref{sec:app}.
\begin{table}[!htbp]
	\renewcommand\arraystretch{1.6}
	\centering
	\caption{The circuit complexity of SIMON32/64}
	\resizebox{\textwidth}{!}{
		\begin{tabular}{ccccccccccccccc}	
			\hline
			\multirow{2}{*}{Round} &\multirow{2}{*}{\#NOT}  & 	\multicolumn{2}{c}{\#$CNOT_{sum}$} & \multicolumn{2}{c}{\#$H_{sum}$}
			& \multirow{2}{*}{\#Toff-S} &\multirow{2}{*}{\#Cliff} & \multirow{2}{*}{\#T} &\multirow{2}{*}{T-depth} &\multirow{2}{*}{\small{Full-depth}} &\multirow{2}{*}{\#qubit}&\multirow{2}{*}{Refer} \\
			\cmidrule(lr){3-4} 
			\cmidrule(lr){5-6} 
			& & \#CNOT & \#Toff-C & \#H  & \#Toff-H  \\ \hline
			32 & 448 & 2816 & 3072 & 0 &1024 & 512 & 7872 & 3584 & - & 946 & 96 & \cite{amm20}\\ 
			32 & 448 & 2816 & 3584 & 0 &1024 & 512 & 8384 & 3584 & 288 & 1024 & 96 & This paper\\     
			19 & 240 & 1568 & 2128 & 0 & 608 & 304 & 4848 & 2128 &171 &608 & 96 & This paper\\     \hline		
		\end{tabular}
	}
	\label{tb:simonimpres}
\end{table}
\subsection{The Circuit Complexity of $\mathcal{QMKS}$ on 19-round SIMON32/64}
\begin{figure}[!htbp]
	\centering
	\includegraphics[width=1\textwidth]{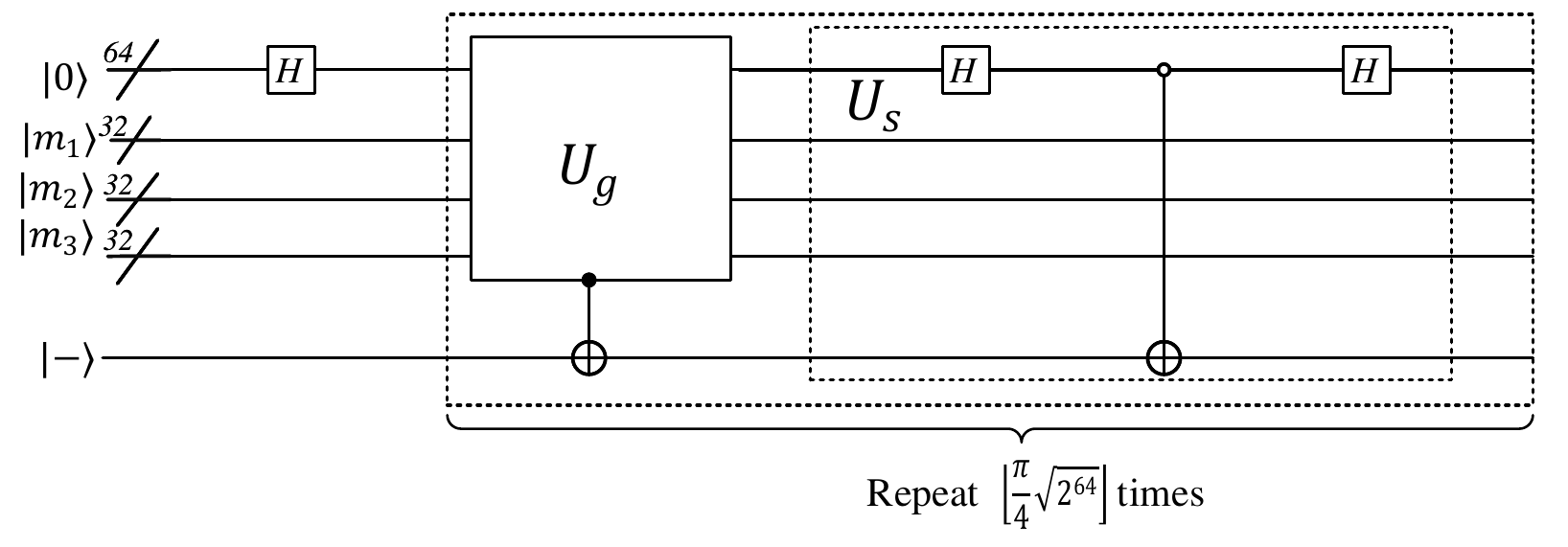}
	\caption{The quantum circuit of $\mathcal{QMKS}$}
	\label{fig:exhtdiag}
\end{figure}

\begin{figure}[!htbp]
	\centering
	\includegraphics[width=1\textwidth]{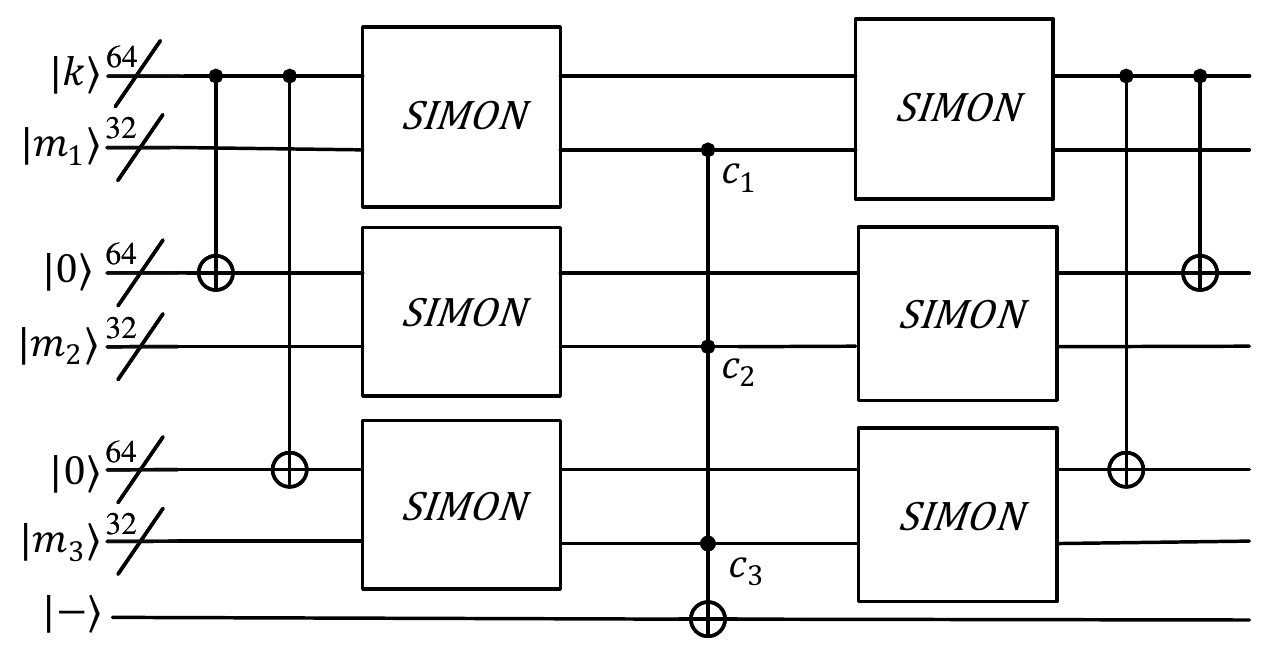}
	\caption{The quantum circuit of $U_g$ in Fig. \ref{fig:exhtdiag}}
	\label{fig:simonexhtora}
\end{figure}

\indent Based on the result in Section \ref{sec:simon}, we reanalyze the quantum circuit complexity of $\mathcal{QMKS}$, whose circuit is shown in Fig. \ref{fig:exhtdiag}. To implement the circuit in Fig. \ref{fig:exhtdiag}, we need to implement the QAA iterator $G=U_sU_g$. The implementation of $U_g$ is in Fig. \ref{fig:simonexhtora}, in which 3 plaintext-ciphertext pairs are chosen for the uniqueness of the solution. The operator $U_s$ consists of two 64-fold Hardmard gates and one 64-fold controlled-NOT gate. Here, we reanalyze the quantum circuit complexity of $\mathcal{QMKS}$ on SIMON32/64 from the following three points. 
\begin{enumerate}
	\item The number of T gates in QAA iterator $G$ is reduced. Anand et al. used the method in \cite{wr16} to implement multi controlled-NOT gates while we use the method shown in Fig. \ref{fig:nfoldCNOT}. In Anand et al.'s circuit, the number of T gates needed to implement a $n$-fold controlled-NOT gate is $32n-84$ while the number is $14n-21$ in our circuit. When $n>4$, the number of T gates in our circuit is smaller than that in the circuit of \cite{amm20}. Therefore, although our circuit uses some more ancillary qubits, the number of T gates is reduced, which makes sense in reducing the complexity of the circuit.
	\item It is enough to perform key expansion in $U_g$ twice, one computation and one uncomputation. In Anand et al.'s estimate, six key expansion processes for six SIMON instances were performed separately in $U_g$, which made the number of NOT gates and CNOT gates were overestimated. There are 448 NOT gates and 1792 CNOT gates during a key expansion process. It's easy to derive that \#NOT=$448\times2=896$. Besides, the CNOT gates come from two key expansion processes and the implementation of six SIMON instances. That is, we can have \#CNOT=$28\times64\times2+32\times32\times6=9728$.
	\item The Clifford gates decomposed by Toffoli gates should be taken into account in gate count estimate. Anand et al. ignored the Clifford gates decomposed by Toffoli gates. The Toffoli gates of the quantum circuit in Fig. \ref{fig:exhtdiag} come from the implementation of SIMON instances and the decomposition of two multi controlled-NOT gates. There are $512\times6=3072$ Toffoli gates in six SIMON instances. Besides, according to the decomposition of Toffoli gate in Fig. \ref{fig:nfoldCNOT}, 96-fold controlled-NOT gate in $U_g$ and 64-fold controlled-NOT gate in $U_s$ can be decomposed into $2\times(96+64)-3=317$ Toffoli gates using 94 ancilla qubits. So we have \#Toff-C=$3389\times7=23723$, \#Toff-H=$3389\times2=6778$. 
	\item The circuit depth estimate result should be more thorough, and the T-depth and Full-depth of QAA iterator $G$ could be reduced further. We decompose the two multi controlled-NOT gates in iterator $G$ into Toffoli gates, and then decompose all Toffoli gates into Clifford gates and T gates. We consider the depth of such circuits that only include Clifford gates and T gates. Although there are six SIMON instances in $U_g$, we only need to consider the depth of two SIMON instances because three SIMON instances are executed in parallel. However, we found that Anand et al. counted the depth of the six SIMON instances into the total depth of $G$ in \cite{amm20}, which overestimated the Full-depth and T-depth of $G$. We estimated that the Toffoli-depth of QAA iterator $G$ is 509, with which we can easily compute T-depth and Full-depth. We can observe that our estimated T-depth and Full-depth are both smaller than those in \cite{amm20}. This is due to the slight modifications we made to the circuit implementation of SIMON32/64. In addition, we didn't ignore the depth of implementing the two multi-control NOT gates, which makes our estimate more accurate and thorough.
\end{enumerate}


\indent Through the above analysis, we present our more accurate estimate results of QAA iterator $G$ in Table \ref{tb:qmkritrres}. To find the master key in the key space $\{0,1\}^{64}$, we need to iterate QAA iterator $G=U_sU_g$ for $\lfloor \frac{\pi}{4}2^{32} \rfloor$ times. From the result in Table \ref{tb:qmkritrres}, we can easily get the quantum circuit complexity of $\mathcal{QMKS}$ on SIMON32/64 in Table \ref{tb:qmkrressum}. In our estimate, the number of Clifford gates is a little higher than that in \cite{amm20} because we consider the number of Clifford gates decomposed by Toffoli gates. Besides, we reduce the number of T gates by adopting the decomposition of multi controlled-NOT gate in Fig. \ref{fig:nfoldCNOT}, which also increases the number of qubits. Also we reduced the T-depth and Full-depth due to our small modifications to the implementation of SIMON32/64. In summary, our estimate result is more accurate and detailed. Through a similar analysis process, we estimated the quantum circuit complexity of the quantum master key exhaustive search attack on SIMON48 and SIMON64 respectively in Table \ref{tb:qmkrressum2} of Appendix \ref{app2:sec2}.

\begin{table}[!htbp]
	\caption{The circuit complexity QAA iterator $G=U_sU_g$ in Fig. \ref{fig:exhtdiag}}
	\renewcommand\arraystretch{1.5}
	\resizebox{\textwidth}{!}{
		\begin{tabular}{cccccccccccccc}		
			\hline
			\multirow{2}{*}{Round} &\multirow{2}{*}{\#NOT}  & 	\multicolumn{2}{c}{\#$CNOT_{sum}$} & \multicolumn{2}{c}{\#$H_{sum}$}&\multirow{2}{*}{\#Toff-S}
			&\multirow{2}{*}{\#Cliff} & \multirow{2}{*}{\#T} 
			&\multirow{2}{*}{T-depth}&\multirow{2}{*}{Full-depth}&\multirow{2}{*}{\#qubit} &\multirow{2}{*}{Refer.} \\
			\cmidrule(lr){3-4} 
			\cmidrule(lr){5-6} 
			& & \#CNOT & \#Toff-C & \#H & \#Toff-H &  \\ \hline
			32 & 2688 & 17152 & 0 & 0 & 0 & 0 & 19840 & 24492  &12288 &27180 &161 & \cite{amm20}\\
			32 & 896 & 9728 & 23723 & 128 & 6778 & 3389 & 44642 & 23723 &1527 & 5318 & 255 & 	This paper\\      
			19 & 480 & 5568 & 14987 & 128 & 4282 & 2141  & 27586 & 14987 &1293 & 4434 & 255 & 	This paper\\    \hline
			
		\end{tabular}
	}
	\label{tb:qmkritrres}
\end{table}
\begin{table}[!htbp]
	\renewcommand\arraystretch{1.6}
	\centering
	\caption{The circuit complexity of $\mathcal{QMKS}$ on SIMON32/64}
	\resizebox{\textwidth}{!}{
		\begin{tabular}{ccccccccccccc}
			\hline
			
			\multirow{2}{*}{Round} &\multirow{2}{*}{\#NOT}  & 	\multicolumn{2}{c}{\#$CNOT_{sum}$} & \multicolumn{2}{c}{\#$H_{sum}$}&\multirow{2}{*}{\#Toff-S} &\multirow{2}{*}{\#Cliff} & \multirow{2}{*}{\#T}
			&\multirow{2}{*}{T-depth}
			&\multirow{2}{*}{Full-depth} &\multirow{2}{*}{\#qubit} &\multirow{2}{*}{Refer.} \\
			
			\cmidrule(lr){3-4} 
			\cmidrule(lr){5-6} 
			& & \#CNOT & \#Toff-C & \#H & \#Toff-H &  \\ \hline
			32 & $2^{43}$ & $1.62 \cdot 2^{45}$ & 0 & 0 & 0 & 0 & $1.35 \cdot 2^{45.5}$ & $1.27 \cdot 2^{46}$ &$1.18\cdot2^{45}$ &$1.05\cdot2^{46.3}$&161 & \cite{amm20}\\ 
			32 & $1.41 \cdot 2^{41}$ & $1.87 \cdot 2^{44}$ & $1.15 \cdot 2^{46}$ & $1.62 \cdot 2^{38}$  & $1.32 \cdot 2^{44}$ & $1.32 \cdot 2^{43}$ & $1.07 \cdot2^{47}$ & $1.15 \cdot 2^{46}$ &$1.15\cdot2^{42}$ &$2^{44}$&255 & This paper\\
			19 & $1.52 \cdot 2^{40}$ & $1.07 \cdot 2^{44}$ & $1.41 \cdot 2^{45}$ & $1.62 \cdot 2^{38}$  &$1.62\cdot 2^{43}$ &$1.62\cdot 2^{42}$ &$1.32 \cdot 2^{46}$ & $1.41 \cdot 2^{45}$ &$2^{42}$ &$1.74\cdot2^{43}$& 255 & This paper\\
			
			\hline
			
		\end{tabular}
	}
	\label{tb:qmkrressum}
\end{table}
	\section{The Quantum Round Key Recovery Attack on 19-round SIMON32/64}
\label{sec:qrkr}
In this section, we describe the quantum round key recovery attack on 19-round SIMON32/64 and give the corresponding quantum circuit as well as its quantum circuit complexity. At first, we recall the classical key recovery attack on 19-round SIMON32/64 in \cite{brv14} where Biryukov et al. present four 13-round differentials with which they recovered the round keys from Round-16 to Round-19. Then we use the four 13-round differentials in \cite{brv14} as our distinguisher and apply QAA algorithm into the two phases of key recovery attack on 19-round SIMON32/64. 
\subsection{The Classical Key Recovery Attack on SIMON 32/64} 
Biryukov et al. proposed the following four 13-round differentials for SIMON32/64.
\begin{equation*}
	\begin{aligned}
		&\mathcal{D}_1:\Delta_{in}^1=(0000,0020),\Delta_{out}^1=(2000,0000)\\
		&\mathcal{D}_2:\Delta_{in}^2=(0000,0040),\Delta_{out}^2=(4000,0000)\\
		&\mathcal{D}_3:\Delta_{in}^3=(0000,0400),\Delta_{out}^3=(0004,0000)\\
		&\mathcal{D}_4:\Delta_{in}^4=(0000,0800),\Delta_{out}^4=(0008,0000)\\
	\end{aligned}
\end{equation*}
To carry out the key recovery attack on 19-round SIMON32/64, two rounds should be added on the top and four rounds on the bottom. Corresponding to each differential, the input truncated differential at the beginning of Rould-1 should be like: 
\begin{equation*}
	\begin{aligned}
		&\Delta{x_1}=(\texttt{00*0 0000 1*00 0000, **00 001* *0*0 0000})\\
		&\Delta{x_2}=(\texttt{0*00 0001 *000 0000, *000 01** 0*00 000*})\\
		&\Delta{x_3}=(\texttt{0001 *000 0000 0*00, 01** 0*00 000* *000})\\
		&\Delta{x_4}=(\texttt{001* 0000 0000 *000, 1**0 *000 00** 0000})\\
	\end{aligned}
\end{equation*}
\indent The process of key recovery process on 19-round SIMON32/64 in \cite{brv14} can be divided into four steps. 
\begin{enumerate}
	\item \emph{Plaintexts Collecting}\label{step1}: Similar to \cite{brv14}, we construct a set $\mathcal{P}$ with $2^{23}$ plaintexts with 9 bits fixed. 
	While different from \cite{brv14}, we just need one right pair.
	By varying some fixed bits of plaintexts in $\mathcal{P}$ and guessing 2 bits of the round key $K^{0}$, we can identify $2^{28.5}$ pairs that satisfy the input difference $\Delta{x_i}$ to Round-3 for each $\mathcal{D}_{i}$ and for each guessed two bits of $K^0$. In total, we can get a set with $2^{30.5}$ plaintext pairs for each $\mathcal{D}_i$ and there must be a right pair in this set.
	\item \emph{Filtering}\label{step2}: $2^{30.5}$ pairs of plaintexts are filtered by verifying the fixed 14 bits of the corresponding difference $\Delta^{18}$.  After filtering, the number of plaintext pairs can be reduced to $2^{30.5-18}=2^{12.5}$ for each differential.
	\item \emph{Partial key guessing}\label{step3}: For each differential, we 	need to recover the following 25 key bits.
	\begin{equation*}
		\begin{aligned}
			&\mathcal{D}^{K}_{1}=\{K^{18},K^{17}[3,5-8,12,14],K^{16}[6]\oplus 	K^{17}[4],K^{16}\oplus K^{17}[2]\}\\
			&\mathcal{D}_2^K=\{K^{18},K^{17}[4,6-9,13,15],K^{16}[7]\oplus 	K^{17}[5],K^{16}[5]\oplus K^{17}[3]\}\\
			&\mathcal{D}_3^K=\{K^{18},K^{17}[8,10-13,1,3],K^{16}[11]\oplus 	K^{17}[9],K^{16}[9]\oplus K^{17}[7]\}\\
			&\mathcal{D}_4^K=\{K^{18},K^{17}[9,11-14,2,4],K^{16}[12]\oplus K^{17}[10],K^{16}[10]\oplus K^{17}[8]\}\\
		\end{aligned}
	\end{equation*}
	The key recovery process of using four differentials is quite similar. So we only describe the key recovery process using $\mathcal{D}_2$. We denote all the key bits in $\mathcal{D}^{K}_{2}$ by $k_1$ and denote the input ciphertext pair by $C=(L^{19},R^{19}),C^{\prime}=((L^{19})^{\prime},(R^{19})^{\prime})$. The keys that satisfy Eq.(\ref{eq:diffequ}) are called candidate keys.
	\begin{equation}
		\label{eq:diffequ}
		D_{k_{1}}^{4}(C)\oplus D_{k_{1}}^{4}(C^{\prime})=\Delta_{out}^2
	\end{equation}
	Eq.(\ref{eq:diffequ}) holds with probability $2^{-14}$, which means there are $2^{25}\times 2^{12.5}/2^{14}=2^{23.5}$ plaintext-key pairs that satisfy Eq.(\ref{eq:diffequ}). In expectation, we can get $2^{23.5}$ candidate keys for $\mathcal{D}^{K}_{2}$. Then we use the other three differentials to carry out the similar key recovery process and can get $2^{23.5}$ candidate keys for $\mathcal{D}^{K}_{1},\mathcal{D}_3^K,\mathcal{D}_4^K$ separately. Because there are some common bits among $\mathcal{D}^{K}_{1},\mathcal{D}_2^K,\mathcal{D}_3^K,\mathcal{D}_4^K$, we can obtain $(2^{23.5})^{4}/(2^{19}\times2^{20}\times2^{22})=2^{33}$ candidate keys for 39 key bits in the last 3 round-keys, i.e. $\mathcal{D}^c=\{K^{18},K^{17}[1-15],K^{16}[6]\oplus K^{17}[4],K^{16}\oplus K^{17}[2],K^{17}[4,6-9,13,15],K^{16}[7]\oplus K^{17}[5],K^{16}[5]\oplus K^{17}[3],\\K^{16}[11]\oplus K^{17}[9],K^{16}[9]\oplus K^{17}[7],K^{16}[12]\oplus K^{17}[10],K^{16}[10]\oplus K^{17}[8]\}$. For simplicity, we denote the 39 key bits by $k_1^{\prime}$. 
	\item \emph{Remaining keys search}\label{step4}: We randomly pick two plaintexts $m_1,m_2$ and get its corresponding ciphertext $c_1,c_2$. We run an exhaustive search on $2^{33}$ candidate keys for 39 key bits in $\mathcal{D}^c$ denoted by $k_1^{\prime}$ and $2^{25}$ remaining 25 key bits denoted by $k_2$ to get the unique and correct key that satisfies $E_{k_1^{\prime}||k_2}(m_1)=c_1	\land E_{k_1^{\prime}||k_2}(m_2)=c_2$. 
\end{enumerate}

\subsection{The Quantum Partial Key Guessing Phase in $\mathcal{QRKR}$ on 19-round SIMON32/64}
\begin{figure}[!htbp]
	\centering
	\includegraphics[width=1\textwidth]{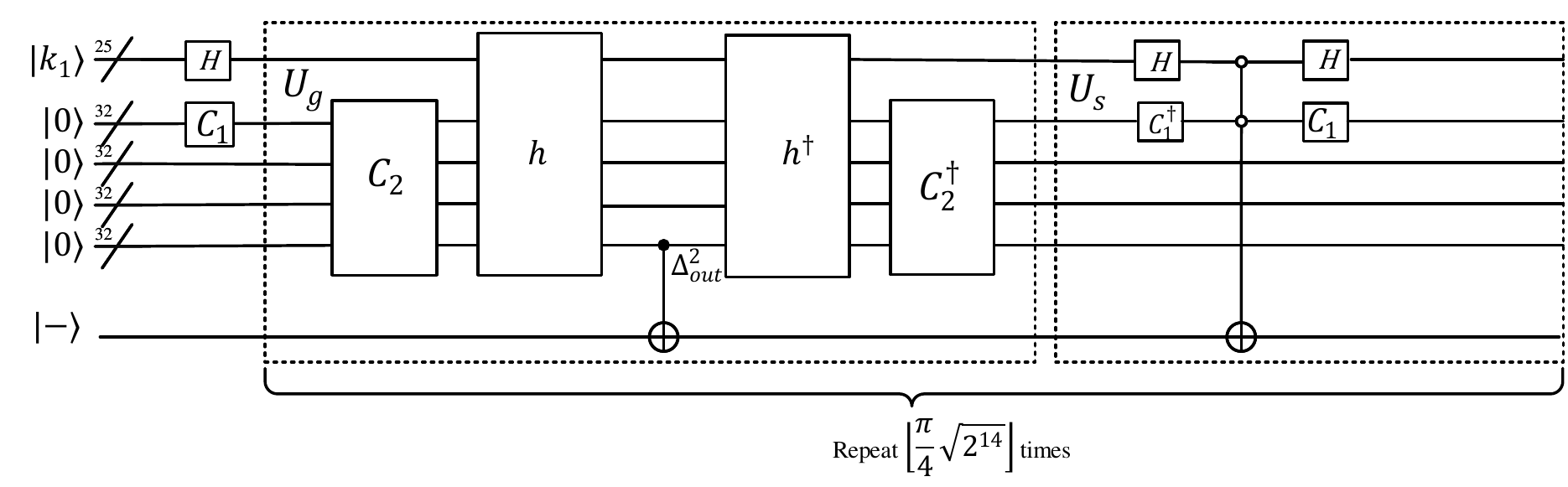}
	\caption{The quantum circuit of partial key guessing using $\mathcal{D}_2$}
	\label{fig:pkgdiag}
\end{figure}

We consider Q1 model as our attack model where both \emph{Plaintexts Collecting} and \emph{Filtering} are classical processes. Thus to design the quantum circuit of quantum key recovery, we only need to regard \emph{Partial key guessing} and \emph{Remaining keys search} as two QAA instances separately. Here we offer the quantum circuit of the first QAA instance and its circuit complexity analysis at first.\\ 
\indent In the quantum partial key guessing phase, four differentials are used to get candidate keys for 39 key bits in $\mathcal{D}_c$. So the QAA instance of this phrase is actually the combination of four sub-QAA instances corresponding to the four processes of partial key guessing using four differentials. The input of every sub-QAA instance is $2^{25}$ partial keys and $2^{12.5}$ plaintext pairs, while the output is a superposition state of $2^{23.5}$ plaintext-key pairs. We need to design the quantum circuit for each sub-QAA instance. Once we have the quantum circuit of one sub-QAA instance using one differential, we can easily design the other three quantum circuits for the other three sub-QAA instances because the four key recovery processes using four differentials are quite similar. Besides, after our analysis, the cost of these four quantum circuits are totally the same. Thus here we only provide the quantum circuit of key recovery process using $\mathcal{D}_2$. \\
\indent Our sub-QAA instance searches the key-plaintext pairs that satisfy Eq.(\ref{eq:diffequ}). The quantum circuit of this sub-QAA instance is in Fig. \ref{fig:pkgdiag}. To achieve our attack, we need to implement two operators $C_1$ and $C_2$ when given classical tuples $(m_i,E(m_i),E(m_i\oplus\Delta{x_2})), i=1,\cdots, 2^{12.5}$. The operator $C_1$ is defined as $C_1\ket{0}\ket{0}\ket{0}=\sum_{i=1}^{2^{12.5}}\ket{m_i}\ket{0}\ket{0}$. And the operator $C_2$ is defined as $C_2\sum_{i=1}^{2^{12.5}}\ket{m_i}\ket{0}\ket{0}=\sum_{i=1}^{2^{12.5}}\ket{m_i}\ket{E(m_i)}\ket{E(m_i\oplus \Delta{x_2})}$.
We suppose the implementation of operator $C_1$ and $C_2$ is efficient so that the cost of operator $C_1$ and $C_2$ can be ignored. To implement the quantum circuit in Fig. \ref{fig:pkgdiag},  $U_g$ and $U_s$ should be implemented separately. The main cost of operator $U_s$ comes from one 57-fold controlled-NOT gate. The main cost of operator $U_g$ comes from the computation of $h$ and one 32-fold controlled-NOT gate. The operator $h$ corresponds to the process of computing $\Delta^{15}$ from given ciphertext pairs, denoted by $(E(m),E(m\oplus\Delta{x_2}))$ and 25 key bits in $\mathcal{D}^K_2$, denoted by $k_1$. 

\indent At first, we define a function $h$ as follows.
\begin{equation*}
	\label{eq:funcf1}
	\begin{aligned}
		h:\{0,1\}^{32} \times \{0,1\}^{25} &\rightarrow\{0,1\}^{32}\\
		(m,k_{1})&\rightarrow D_{k_{1}}^{4}(E(m))\oplus D_{k_{1}}^{4}(E(m\oplus \Delta {x_2}))
	\end{aligned}
\end{equation*}
Then based on $h$ we define a Boolean function $g$ as follows.
\begin{eqnarray*}
	g(m,k_{1})=
	\begin{cases}
		1, \text{if} \; h(m,k_{1})=\Delta_{out}^2\\
		0, \text{if} \; h(m,k_{1})\neq \Delta_{out}^2\\
	\end{cases}
\end{eqnarray*}
Naturally, the operator $U_g$ is defined as follows:
\begin{eqnarray*}
	U_g\ket{k_1}\ket{m}\ket{0}\ket{0}\ket{0}=
	\begin{cases}
		\ket{k_1}\ket{m}\ket{0}\ket{0}\ket{0}, 	\text{if} \; g(m,k_{1})=0\\
		-\ket{k_1}\ket{m}\ket{0}\ket{0}\ket{0}, 	\text{if} \; g(m,k_{1})=1\\
	\end{cases}
\end{eqnarray*}

\begin{figure}[!htbp]
	\centering
	\begin{subfigure}[b]{1\textwidth}
		\includegraphics[width=\textwidth]{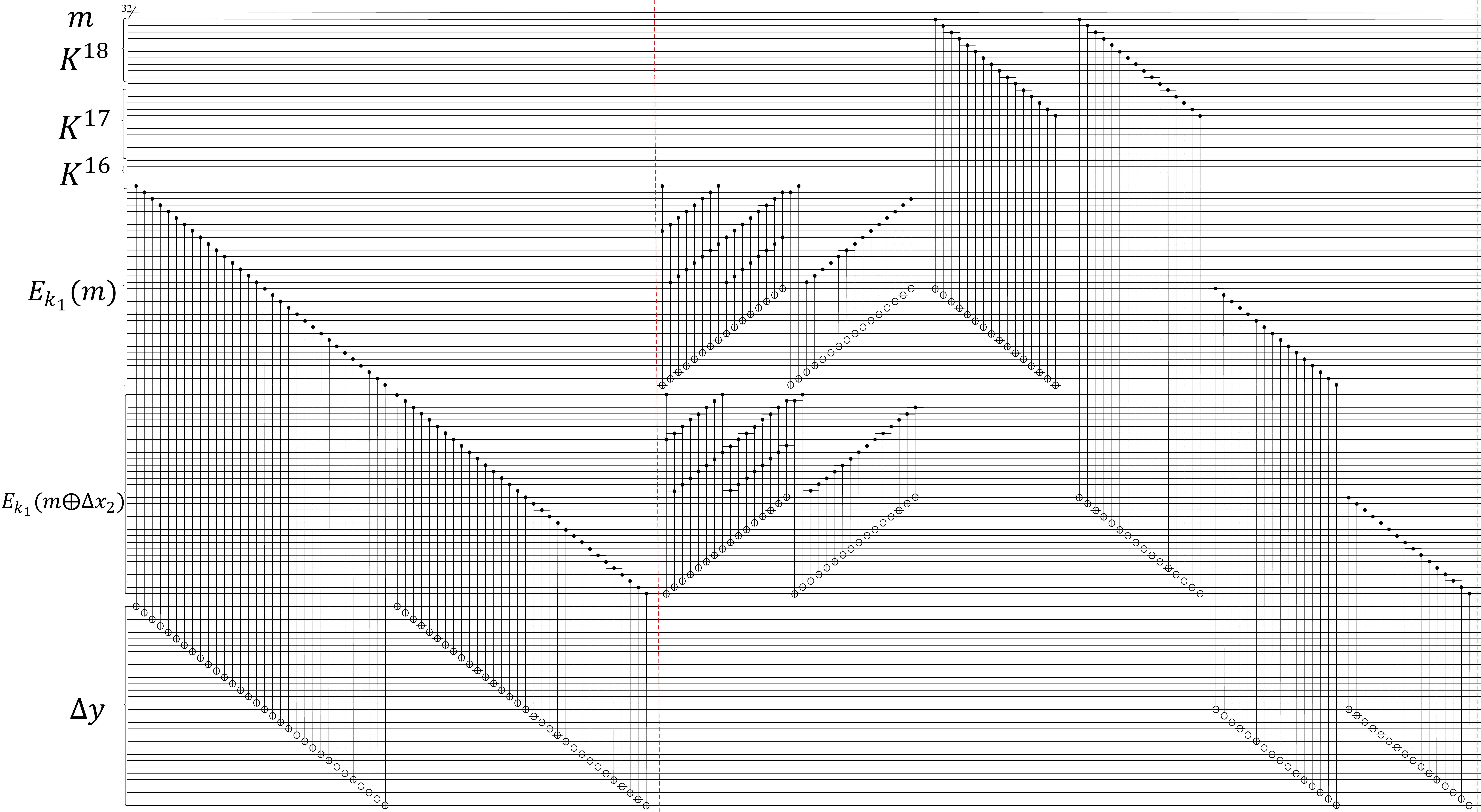}
		\caption{The quantum circuit of computing  $\Delta^{18}$ from input ciphertext pairs}
		\label{fig:hcircuit1}
	\end{subfigure}
	\\
	\begin{subfigure}[b]{1\textwidth}
		\includegraphics[width=\textwidth]{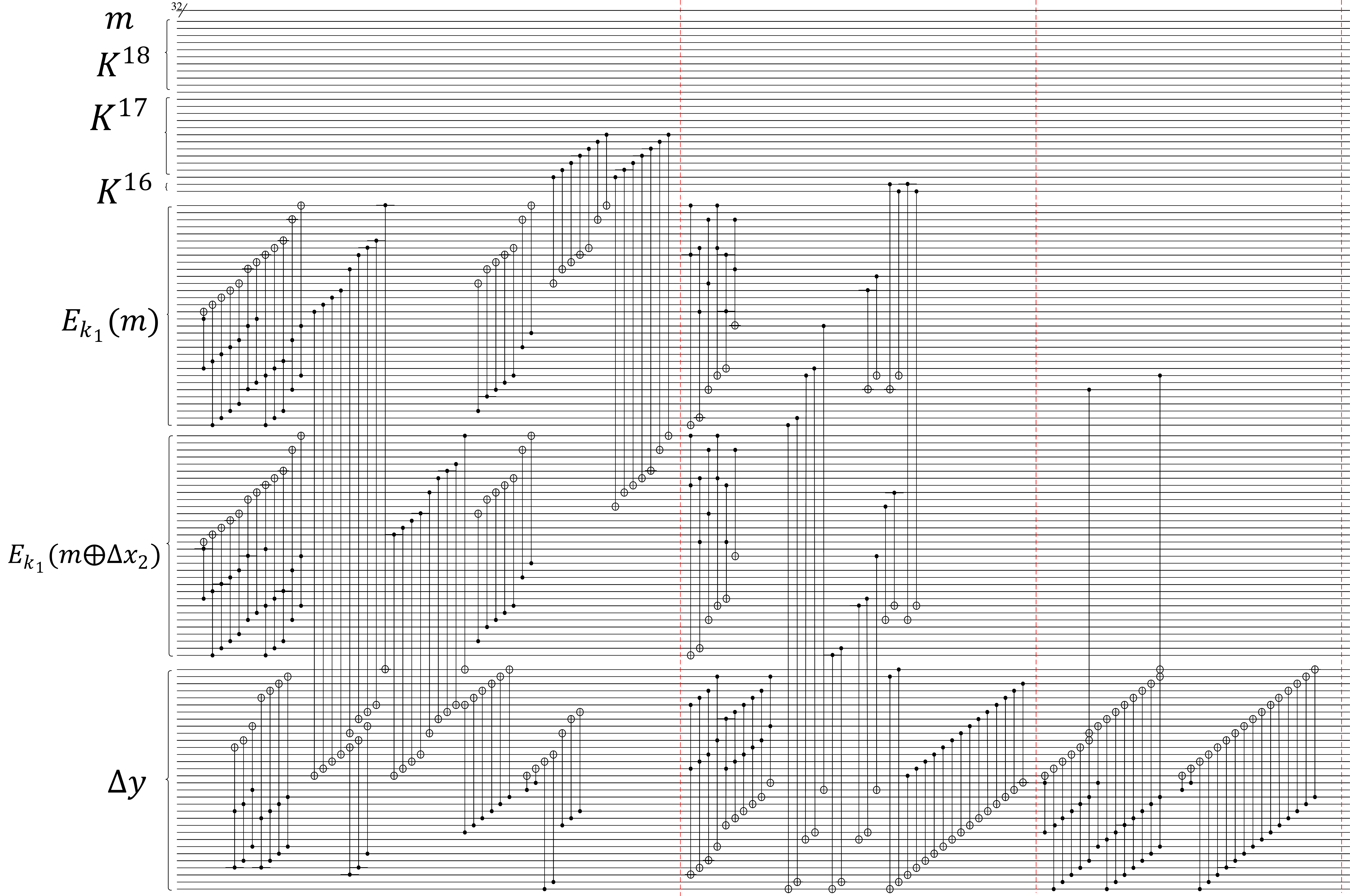}
		\caption{The quantum circuit of computing  $\Delta^{15}$ from $\Delta^{18}$}
		\label{fig:hcircuit2}
	\end{subfigure}
	
	\caption{The quantum circuit of function $h$($\Delta y=E_{k_1}(m)\oplus E_{k_1}(m\oplus \Delta x_2)$)}
	\label{fig:hcircuit}
\end{figure}

\indent The quantum circuit of function $h$ should be designed in detail, before which the computation process of $h$ needs to be sorted out clearly. We denote the input ciphertext pair by $E(m)=(L^{19},R^{19}),E(m\oplus \Delta {x_2})=((L^{19})^{\prime},(R^{19})^{\prime})$. The computation process to get $\Delta^{15}$ using $\mathcal{D}_2$ is as follows:
\begin{enumerate}
	\item From the given ciphertext pair, we can easily get $\Delta ^{19}=(L^{19}\oplus 	(L^{19})^{\prime},R^{19}\oplus (R^{19})^{\prime})$.
	\item According to the guess of $K^{18}$, we can easily get $(L^{18},R^{18})=F(L^{19},R^{19}),((L^{18})^{\prime},\\(R^{18})^{\prime})=F((L^{19})^{\prime},(R^{19})^{\prime}),\Delta ^{18}=(L^{18}\oplus (L^{18})^{\prime},R^{18}\oplus (R^{18})^{\prime})$.
	\item On one hand, we compute $\Delta ^{17}$ via Eq. (\ref{comptd17}).
	\begin{equation}
		\label{comptd17}
		\resizebox{0.95\hsize}{!}{
			$\left\{
			\begin{array}{ll}
				\Delta L^{17}=\Delta R^{18}, \\
				\Delta R^{17}[i]=\Delta And^{17}[i]\oplus \Delta Rot^{17}[i]\oplus \Delta L^{18}[i],  	&i=\{0,1,2,3,6,8,9,10,15\}\\
				\Delta R^{17}[i]=\Delta L^{17}[(i-1) \% 16] \& \Delta L^{17}[(i-8) \% 16]\oplus \Delta L^{17}[(i-2) \% 16]\oplus \Delta L^{18}[i],  	&i\in \{0,1,\cdots,15\}\setminus \{0,1,2,3,4,6,8,9,15\}\\				
			\end{array}
			\right.$
		}
	\end{equation}
	
	On the other hand, for the necessity of computing $\Delta ^{16}$, we should use guessed key bits $K^{17}[4,6-9,13,15]$ to compute $R^{17}[4,6-9,13,15],(R^{17})^{\prime}[4,6-9,13,15]$ and use guessed key bits $K^{17}[3]\oplus K^{16}[5],K^{17}[5]\oplus K^{16}[7]$ to compute $R^{17}[i]\oplus K^{16}[i+2],(R^{17})^{\prime}[i]\oplus K^{16}[i+2],i\in\{3,5\}$.
	
	\begin{equation}
		\label{comptv17}
		\resizebox{0.95\hsize}{!}{
			$\left\{
			\begin{array}{ll}
				R^{17}[i]= L^{17}[(i-1) \% 16]\& L^{17}[(i-8) \% 16] \oplus L^{17}[(i-2) \% 16] \oplus L^{18}[i]\oplus K^{17}[i],  	&i\in \{4,6-9,13,15\}\\
				(R^{17})^{\prime}[i]= (L^{17})^{\prime}[(i-1) \% 16]\& (L^{17})^{\prime}[(i-8) \% 16] \oplus (L^{17})^{\prime}[(i-2) \% 16] \oplus (L^{18})^{\prime}[i]\oplus K^{17}[i],  	&i\in \{4,6-9,13,15\}\\
				
				R^{17}[i]\oplus K^{16}[i+2]=L^{17}[(i-1)\%16]\&L^{17}[(i-8)\%16]\oplus L^{17}[(i-2)\%16]\oplus K^{17}[i]\oplus K^{16}[i+2],&i\in\{3,5\}\\
				(R^{17})^{\prime}[i]\oplus K^{16}[i+2]=(L^{17})^{\prime}[(i-1)\%16]\&(L^{17})^{\prime}[(i-8)\%16]\oplus (L^{17})^{\prime}[(i-2)\%16]\oplus K^{17}[i]\oplus K^{16}[i+2],&i\in\{3,5\}\\						
			\end{array}
			\right.$
		}
	\end{equation}
	
	\item On one hand, we compute $\Delta ^{16}$ according to Eq. (\ref{comptd16}).
	\begin{equation}
		\label{comptd16}
		\resizebox{0.95\hsize}{!}{
			$\left\{
			\begin{array}{ll}
				\Delta L^{16}=\Delta R^{17}, \\
				\Delta R^{16}[i]=\Delta And^{16}[i] \oplus \Delta Rot^{16}[i] \oplus \Delta L^{17}[i],  &i\in\{0,1,7,8,14\}\\
				\Delta R^{16}[i]=\Delta L^{16}[(i-1) \% 16] \& \Delta L^{16}[(i-8) \% 16]\oplus \Delta L^{16}[(i-2) \% 16]\oplus \Delta L^{17}[i], &i\in\{0,\cdots,15\}\setminus\{0,1,7,8,14\}\\
			\end{array}
			\right.$
		}
	\end{equation}
	
	On the other hand, for the necessity of computing $\Delta^{15}$, we need to compute $R^{16}[5,7]$ and $(R^{16})^{'}[5,7]$.
	\begin{equation}
		\label{comptv16}
		\resizebox{0.95\hsize}{!}{
			$\left\{
			\begin{array}{ll}
				R^{16}[i]= L^{16}[(i-1) \% 16]\& L^{16}[(i-8) \% 16] \oplus L^{17}[i] \oplus (L^{16}[(i-2) \% 16]\oplus K^{16}[i]) ,  	&i\in \{5,7\}\\
				(R^{16})^{\prime}[i]= (L^{16})^{\prime}[(i-1) \% 16]\& (L^{16})^{\prime}[(i-8) \% 16] \oplus (L^{17})^{\prime}[i]\oplus ((L^{16})^{\prime}[(i-2) \% 16] \oplus K^{17}[i]),  	&i\in \{5,7\}\\
				
			\end{array}
			\right.$
		}
	\end{equation}
	\item At last, we compute $\Delta ^{15}$ according to Eq. (\ref{comptd15}).
	\begin{equation}
		\label{comptd15}
		\resizebox{0.95\hsize}{!}{
			$\left\{
			\begin{array}{ll}
				\Delta L^{15}=\Delta R^{16},\\
				\Delta R^{15}[6]=R^{16}[5]\oplus \Delta L^{15}[3],\\
				\Delta R^{15}[15]=R^{16}[7]\oplus \Delta L^{15}[13],\\
				\Delta R^{15}[i]=\Delta L^{16}[(i-1) \% 16] \& \Delta L^{16}[(i-8) \% 16]\oplus \Delta L^{16}[(i-2) \% 16]\oplus \Delta L^{17}[i], &i\in\{0,\cdots,15\}\setminus\{6,15\}\\
			\end{array}
			\right.$
		}
	\end{equation}
	
\end{enumerate}
\indent According to the above process, we provide our quantum circuit of $h$ in Fig. \ref{fig:hcircuit}. After a simple analysis of the circuit, we can easily get there are 244 CNOT gates and 103 Toffoli gates in the implementation of $h$ with Full-depth 124 and T-depth 33.

Having the quantum circuit of $h$, we could easily estimate the circuit complexity of quantum partial key guessing using differential $\mathcal{D}_2$ in Table \ref{tb:qrkrstep3res}. Following the same process, we can easily design the quantum circuit of the other three sub-QAA instances using $\mathcal{D}_1,\mathcal{D}_3,\mathcal{D}_4$ separately. And the circuit complexity of the other three sub-QAA instances can also be seen in Table \ref{tb:qrkrstep3res}. Similarly, based on the 15-round differential path of SIMON48 and the 21-round differential path of SIMON64 given by Biryukov in Table 5 in \cite{brv14}, we also analyzed the circuit complexity of quantum partial key guessing on 19-round SIMON48 and 26-round SIMON64 respectively and listed the result in Table \ref{tb:qrkrstep3res2} of Section \ref{sec:app}.\\

\begin{table}[!htbp]	
	\renewcommand\arraystretch{1.5}
	\centering
	\caption{The circuit complexity of quantum partial key guessing using $\mathcal{D}_i(i=1,2,3,4)$ in $\mathcal{QRKR}$}
	\resizebox{\textwidth}{!}{
		\begin{tabular}{cccccccccccc}
			
			\hline
			\multirow{2}{*}{\#iter} &\multirow{2}{*}{\#NOT}  & \multicolumn{2}{c}{\#$CNOT_{sum}$} & \multicolumn{2}{c}{\#$H_{sum}$}&\multirow{2}{*}{\#Toff-S} &\multirow{2}{*}{\#Cliff} & \multirow{2}{*}{\#T} &\multirow{2}{*}{T-depth}&\multirow{2}{*}{Full-depth}&\multirow{2}{*}{\#qubit} \\
			
			\cmidrule(lr){3-4} 
			\cmidrule(lr){5-6} 
			& & \#CNOT & \#Toff-C & \#H & \#Toff-H &  \\ \hline
			1 & 0 & 488 & 2667 & 50 & 762 &381 & 4348 & 2667 &591 &2000 & 209\\    
			
			$\lfloor \frac{\pi}{4}\sqrt{2^{14}}\rfloor$ & 0 & $1.62 \cdot 2^{15}$  &$ 2^{18}$  &$1.23 \cdot 2^{12}$  &$1.15 \cdot 2^{16}$ &$1.15 \cdot 2^{15}$  & $1.62 \cdot 2^{18}$  & $1.52\cdot 2^{18}$ & $1.74 \cdot 2^{15}$ &  $1.52 \cdot 2^{17}$ & 209\\
			\hline
			
		\end{tabular}
	}
	\label{tb:qrkrstep3res}
\end{table}
\indent At last, we describe our method of generating candidate keys. The four sub-QAA instances in the first QAA instance all output a superposition state of $2^{23.5}$ plaintext-key pair that satisfies Eq.(\ref{eq:diffequ}) among $2^{12.5}$ plaintext pairs and $2^{25}$ partial keys after $\lfloor\frac{\pi}{4}\sqrt{2^{14}}\rfloor$ iterations. To get candidate keys, we measure the key register many times. The probability of measuring the right partial key is $2^{-23.5}$. That is, we expect that we can get the right partial key after running this sub-QAA instance for $2^{23.5}$ times. And in expectation, we can get $2^{23.5}[1-(1-\frac{1}{2^{23.5}})^{2^{23.5}}]\approx2^{22.8}$ different candidate keys for 25 key bits in $\mathcal{D}^K_{2}$ from $2^{23.5}$ measurements. After combining the results of the other three sub-QAA instances, we can get $(2^{22.8})^{4}/(2^{19}\times2^{20}\times2^{22})=2^{30.2}$ candidate keys for 39 key bits in $\mathcal{D}^c$. Despite that the cost of the process is a little high, it was hard to find other more efficient ways to get all candidate keys. Actually, Kaplen et al. also adopted a similar method to generate all candidate keys by measuring the key register many times in \cite{kap16a}. However, in their method, they ensured that the newly gotten candidate key was different from the ones gotten before by excluding the keys that had been gotten in the QAA oracle. To implement their method using a quantum circuit, a sequence of multi controlled-NOT gates need to be added in QAA oracle. That is, for every run, we need to design a new quantum circuit, which would greatly increase the quantum resources. Besides, the number of iteration increases with the increase of the number of elements needed to be excluded, which makes their encryption complexity also high. In our method, despite that we need to measure many times, we do not need to design a new quantum circuit in each run, which saves quantum resources.

\begin{remark}
	We consider a practical model, Q1 model. In Fig. \ref{fig:pkgdiag}, the operator $C_1$ achieves the process of preparing a superposition of $2^{12.5}$ classical plaintexts $m_i,i=1,2,\cdots,2^{12.5}$. And the operator $C_2$ achieves the process of preparing a superposition of $2^{12.5}$ classical tuples $(m_i,E(m_i),E(m_i\oplus \Delta{x}_2))$. Actually, it's not known whether there exists such efficient operators that could achieve such transformation, the difficulty of which is equal to preparing the superposition of random states. The choice of classical tuples may influence the efficiency of operators $C_1$ and $C_2$. If there are structures in the classical tuples, it may be efficient to get the target superposition state. 
\end{remark}
\subsection{The Quantum Remaining Keys Search Phase in $\mathcal{QRKR}$ on 19-round SIMON32/64}
\begin{figure}[!htbp]
	\centering
	\includegraphics[width=1\textwidth]{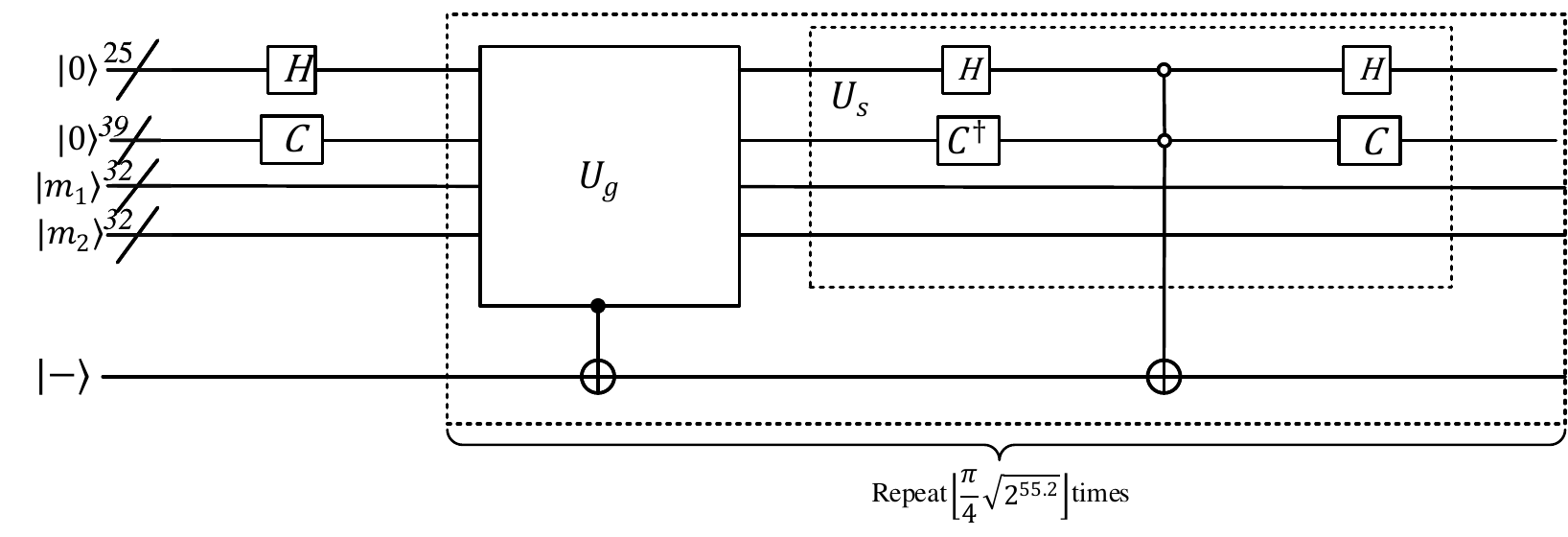}
	\caption{The quantum circuit of remaining keys search phase in $\mathcal{QRKR}$}
	\label{fig:exht2}
\end{figure}

\begin{figure}[!htbp]
	\centering
	\includegraphics[width=1\textwidth]{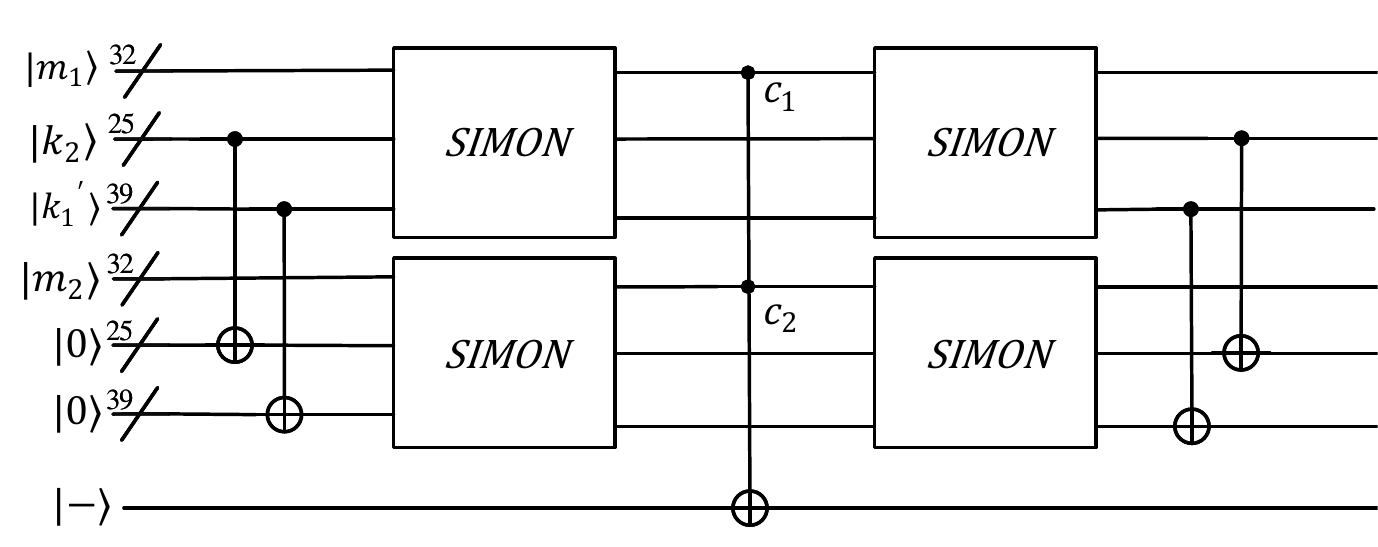}
	\caption{The quantum circuit of $U_g$ in Figure \ref{fig:exht2}}
	\label{fig:exht}
\end{figure}

To complete the attack $\mathcal{QRKR}$, we should use the candidate keys gotten from the first phase to carry out the quantum remaining keys search phase. Here, we offer the quantum circuit of this phase(corresponding to Step \ref{step4}) and analyze its quantum circuit complexity. \\
\indent We define another QAA instance in search space of $2^{30.2}$ candidate keys for 39 key bits in $\mathcal{D}^c$ denoted by $k_1^{\prime}$ and $2^{25}$ remaining 25 key bits denoted by $k_2$. According to \cite{jnr20}, we need to choose two plaintexts $m_1,m_2$ and get their corresponding ciphertexts $c_1,c_2$ in QAA oracle to ensure the uniqueness of the solution. The quantum circuit of this QAA instance is in Fig. \ref{fig:exht2}. The $C$ operator is a creation operator, which creates the superposition state of $2^{30.2}$ candidate keys for 39 key bits in $\mathcal{D}^c$ from the all-zero state, which is defined as $C\ket{0}=\sum_{i=1}^{i=2^{30.2}}\ket{(k_1^{\prime})^i}$. As previously assumed, we also assume that this process is efficient so that the cost of operator $C$ could be ignored. Then, we need to implement the quantum circuit of $U_g$ and $U_s$ separately. The main cost of $U_s$ is one 64-fold controlled-NOT gate. The main cost of $U_g$ comes from four SIMON instances, and the circuit of $U_g$ is shown in Fig. \ref{fig:exht}.

\indent We define a function $h$ as follows, which corresponds to the encryption process of $m_1,m_2$ with given $k_1^{\prime}||k_2$.
\begin{equation*}
	\begin{aligned}
		h:\{0,1\}^{39}\times\{0,1\}^{25} &\rightarrow\{0,1\}^{32}\times\{0,1\}^{32}\\
		(k_{1}^{\prime },k_{2})&\rightarrow (E_{k_{1}^{\prime}||k_{2}}(m_1),E_{k_{1}^{\prime}||k_{2}}(m_2))
	\end{aligned}
\end{equation*}
Then based on $h$, we define a Boolean function $g$.

\begin{eqnarray*}
	g(k_1^{\prime},k_{2})=
	\begin{cases}
		1, \text{if} \; h(k_1^{\prime},k_{2})=(c_1,c_2)\\
		0, \text{if} \; h(k_1^{\prime},k_{2})\neq (c_1,c_2)\\
	\end{cases}
\end{eqnarray*}

\noindent Naturally, the operator $U_g$ is defined as follows:
\begin{eqnarray*}
	U_g\ket{k_1^{\prime}}\ket{k_2}\ket{0}\ket{0}=
	\begin{cases}
		\ket{k_1^{\prime}}\ket{k_2}\ket{0}\ket{0}, 	\text{if} \; g(k_{1}^{\prime},k_{2})=0\\
		-\ket{k_1^{\prime}}\ket{k_2}\ket{0}\ket{0}, \text{if} \; g(k_{1}^{\prime},k_{2})=1\\
	\end{cases}
\end{eqnarray*}

\indent To find the unique and correct round key, the QAA iterator $G=U_sU_g$ should be iterated $\lfloor  \frac{\pi}{4}\sqrt{2^{55.2}}\rfloor$ times. We can easily deduce the circuit complexity of the quantum remaining keys search phase in Table \ref{tb:qrkrstep4res}. Similarly, based on the differential path of SIMON48 and SIMON64 given by Biryukov in Table 5 in \cite{brv14}, we analyzed the circuit complexity of quantum remaining keys on 19-round SIMON and 26-round SIMON64 respectively shown in Table \ref{tb:qrkrstep4res2} of Appendix \ref{sec:app}.

\begin{table}[!htbp]
	\renewcommand\arraystretch{1.5}
	\centering
	\caption{The circuit complexity of quantum remaining keys search}
	\resizebox{\textwidth}{!}{
		\begin{tabular}{cccccccccccc}
			\hline
			\multirow{2}{*}{\#iter} &\multirow{2}{*}{\#NOT}  & \multicolumn{2}{c}{\#$CNOT_{sum}$} & 	\multicolumn{2}{c}{\#$H_{sum}$} &\multirow{2}{*}{\#Toff-S} &\multirow{2}{*}{\#Cliff} & \multirow{2}{*}{\#T}
			& \multirow{2}{*}{T-depth}& \multirow{2}{*}{Full-depth} &\multirow{2}{*}{\#qubit} \\ 
			
			\cmidrule(lr){3-4} 
			\cmidrule(lr){5-6} 
			& & \#CNOT & \#Toff-C & \#H & \#Toff-H &  \\ \hline
			1 & 480 & 3392 & 10262 & 78 & 2932 &1466 & 18610 & 10262 &1092 &3718 & 191\\
			
			$\lfloor \frac{\pi}{4}\sqrt{2^{55.2}}\rfloor$ & $1.15 \cdot 2^{36}$ 
			& $2^{39}$ & $1.52\cdot2^{40}$  &$1.41 \cdot 2^{33}$  &$1.74 \cdot 2^{38}$ &$1.74 \cdot 2^{37}$ & 	$1.32 \cdot 2^{41}$& $1.52\cdot2^{40}$ & $1.23\cdot2^{37}$ & $1.07\cdot2^{39}$ & 191 \\\hline
			
		\end{tabular}
	}
	\label{tb:qrkrstep4res}
\end{table}
	\section{The Complexity Analysis}\label{sec:complexity}
In this section, we compare the complexity of $\mathcal{QMKS}$ on 19-round SIMON32/64 and $\mathcal{QRKR}$ on 19-round SIMON32/64 in terms of encryption complexity and circuit complexity separately.

\subsection{Encryption Complexity Comparison}

\indent In $\mathcal{QMKS}$ on 19-round SIMON32/64, to recover the master key, we need to carry out $$\lfloor\frac{\pi}{4}2^{32}\rfloor\times\frac{19}{32}\times2\approx 2^{32.6}$$ 
encryptions. Despite that there are 6 SIMON instances, 2 encryptions are needed only in one iteration due to parallel. \\
\indent In $\mathcal{QRKR}$ on 19-round SIMON32/64, to recover round key $$4\times2^{23.5}\times\lfloor\frac{\pi}{4}2^{\sqrt{14}}\rfloor\times\frac{4}{19}\times2 + \lfloor\frac{\pi}{4}2^{\sqrt{55.2}}\rfloor\times 2 \approx 2^{31.1}$$
encryptions are needed. In the first term, 4 represents four sub-QAA instances using four differentials, and $\frac{4}{19}$ represents the complexity of 4-round decryption. In the second term, 2 represents 2 processes of SIMON encryption. \\
\indent On the whole, the encryption complexity of $\mathcal{QRKR}$ on 19-round SIMON32/64 is slightly lower than $\mathcal{QMKS}$ on 19-round SIMON32/64. The main encryption complexity comes from generating candidate keys process in the quantum partial key guessing phase. As a result, if the complexity of quantum partial key guessing could be reduced further, $\mathcal{QRKR}$ on 19-round SIMON32/64 could achieve much lower encryption complexity. \\

\subsection{Quantum Circuit Complexity Comparison}		

\indent The QAA instance of quantum partial key guessing phase consists of four sub-QAA instances. So multiplying the gate count in Table \ref{tb:qrkrstep3res} by 4, we can get the gate count of this QAA instance in the second line of Table \ref{tb:rescomp}. However, when computing the T-depth and Full-depth of this QAA instance, we should multiply the T-depth and Full-depth in Table \ref{tb:qrkrstep3res} by the running time $2^{23.5}$. Then the circuit complexity of the quantum partial key guessing phase can be listed in the third line of Table \ref{tb:rescomp}. \\
\indent From Table \ref{tb:rescomp}, we can observe that the quantum gates count of the first phase in $\mathcal{QRKR}$ on 19-round SIMON32/64 is far lower than that of the second phase so that it can be omitted. However, for the reason that many running times are needed for generating candidate keys in the first phase, the gap of T-depth and Full-depth between the first QAA instance and the second QAA instance is small. After comparing the overall circuit complexity of $\mathcal{QRKR}$ on 19-round SIMON32/64 with that of $\mathcal{QMKS}$ on 19-round SIMON32/64 comprehensively, we conclude that the circuit complexity of $\mathcal{QRKR}$ on 19-round SIMON32/64 is lower than that of $\mathcal{QMKS}$ on 19-round SIMON32/64. Besides, we make a similar comparison between quantum round key recovery attack and quantum master key exhaustive search on 19-round SIMON48 and 26-round SIMON64 respectively, shown in Table \ref{tb:comp}.\\ 

\begin{table}[!htbp]
	\renewcommand\arraystretch{1.5}
	\centering
	\caption{The circuit complexity comparison between $\mathcal{QMKS}$ on on 19-round SIMON32/64 and $\mathcal{QRKR}$ on 19-round SIMON32/64}
	\resizebox{\textwidth}{!}{
		\begin{tabular}{cccccccccccc}
			\hline
			\multirow{2}{*}{Algorithm} &\multirow{2}{*}{\#NOT}  & \multicolumn{2}{c}{\#$CNOT_{sum}$} 	& \multicolumn{2}{c}{\#$H_{sum}$}&\multirow{2}{*}{\#Toff-S} &\multirow{2}{*}{\#Cliff} & \multirow{2}{*}{\#T} &\multirow{2}{*}{T-depth}&\multirow{2}{*}{Full-depth} &\multirow{2}{*}{\#qubit} \\
			
			\cmidrule(lr){3-4} 
			\cmidrule(lr){5-6} 
			& & \#CNOT & \#Toff-C & \#H & \#Toff-H &  \\ \hline
			
			$\mathcal{QMKS}$ & $1.52 \cdot 2^{40}$ & $1.07 \cdot 2^{44}$ & $1.41 \cdot 2^{45}$ & $1.62 \cdot 2^{38}$  &$1.62\cdot 2^{43}$ &$1.62\cdot 2^{42}$  &$1.32 \cdot 2^{46}$ & $1.41 \cdot 2^{45}$ & $2^{42}$ & $1.74 \cdot 2^{43}$  &255\\      
			
			\multirow{2}{*}{$\mathcal{QRKR}$ } & 0 & $1.62 \cdot 2^{17}$  &$ 2^{20}$  &$1.23 \cdot 2^{14}$  &$1.15 \cdot 2^{18}$ &$1.15 \cdot 2^{17}$  & $1.62 \cdot 2^{20}$  & $1.52\cdot 2^{20}$ & $1.23 \cdot 2^{39}$ &  $1.07 \cdot 2^{41}$ & 209 \\
			& $1.15 \cdot 2^{36}$ 
			& $2^{39}$ & $1.52\cdot2^{40}$  &$1.41 \cdot 2^{33}$  &$1.74 \cdot 2^{38}$  &$1.74 \cdot 2^{37}$ & 	$1.15 \cdot 2^{41}$& $1.52\cdot2^{40}$ &$1.23 \cdot 2^{37}$ &$1.07 \cdot 2^{39}$  & 191\\\hline
		\end{tabular}
	}
	\label{tb:rescomp}
\end{table}

\indent In summary, we gain a quantum dedicated attack that has lower encryption complexity and quantum circuit complexity than the quantum generic attacks on SIMON32/64. However, we find it's not a big complexity gap between our attack and exhaustive search in quantum setting due to the big complexity of generating candidate keys. 
	\section{Conclusion}
\label{sec:conclusion}
In this paper, we studied the quantum key recovery attack on SIMON block cipher using QAA algorithm in Q1 model. We reanalyzed the quantum circuit complexity of the quantum generic attacks on SIMON32, SIMON48 and SIMON64 and firstly provided quantum dedicated attacks on these SIMON variants. And our work studied quantum dedicated attack on SIMON block cipher from the perspective of quantum circuit complexity for the first time, which can provide a research basis for performing real attacks on quantum computers in the future. \\
\indent On one hand, we gave more accurate and thorough circuit complexity analysis results of the quantum circuit complexity of quantum master key exhaustive search on SIMON32, SIMON48 and SIMON64 than the results in \cite{amm20}. We considered the number of Clifford gates more comprehensively and reduced the number of T gates. And we reduced the T-depth and Full-depth via small modifications to the circuit. On the other hand, we analyzed the circuit complexity of quantum round key recovery attacks on 19-round SIMON32, 19-round SIMON48 and 26-round SIMON64. We took the attack on 19-round SIMON32/64 as an example and designed its circuit in detail. The two phases of key recovery attack on 19-round SIMON32/64 can be regarded as two QAA instances separately, and the first QAA instance is composed of four sub-QAA instances corresponding to the four processes of using four differentials for key recovery. After our analysis, the encryption complexity and quantum circuit complexity of quantum round key recovery attacks on 19-round SIMON32, 19-round SIMON48 and 26-round SIMON64/128 are both lower than those of quantum master key exhaustive search on these SIMON variants. However, we used the method of measuring many times to generate all the candidate keys and failed to find a better way to generate candidate keys, which is the bottleneck of reducing complexity. \\
\indent In the following work, we may try to combine other key recovery techniques with our quantum dedicated attack, such as the dynamic key-guessing techniques proposed by Wang et al. \cite{wn18}. Besides, more efforts should be made to study how to reduce the complexity of generating candidate keys. Further, we could investigate the physical feasibility of our attack by considering the decoherence time of quantum computers and the time of CNOT operation because the two-qubit operation takes a longer time than single-qubit operations.
	\begin{subappendices}
		\renewcommand{\thesection}{\Alph{section}}%
		
		\section{The Complexity Analysis of Quantum Key Recovery Attack on 19-round SIMON48 and 26-round SIMON64}
\label{sec:app}
In this Appendix, we list the circuit complexity of quantum key recovery attacks on  19-round SIMON48 and 26-round SIMON64. Firstly, through the analysis process similar to Section \ref{sec:qmks}, we respectively analyzed the quantum circuit complexity of the quantum master key search on these SIMON variants. Then, based on the key recovery attacks on SIMON48 and SIMON96 given by Biryukov et al. in the Appendix of \cite{brv14}, we respectively analyzed the circuit complexity of the quantum round key recovery attacks on these SIMON variants. At last, we compare the encryption complexity and circuit complexity of quantum master key exhaustive search and quantum round key recovery attack on 19-round SIMON32, 19-round SIMON48 and 26-round SIMON64.

\subsection{The Quantum Circuit Complexity of SIMON48 and SIMON64}
\label{app2:sec1}
\begin{table}[!htbp]
	\renewcommand\arraystretch{1.5}
	\centering
	\caption{The quantum circuit complexity of SIMON48 and SIMON64}
	\resizebox{\textwidth}{!}{
		\begin{tabular}{cccccccccccccccc}	
			\hline
			\multirow{2}{*}{Cipher} &\multirow{2}{*}{Round} &\multirow{2}{*}{\#NOT}  & 	\multicolumn{2}{c}{\#$CNOT_{sum}$} & \multicolumn{2}{c}{\#$H_{sum}$}
			& \multirow{2}{*}{\#Toff-S} &\multirow{2}{*}{\#Cliff} & \multirow{2}{*}{\#T} &\multirow{2}{*}{T-depth} &\multirow{2}{*}{\small{Full-depth}} &\multirow{2}{*}{\#qubit}\\
			\cmidrule(lr){4-5} 
			\cmidrule(lr){6-7} 
			&  & & \#CNOT & \#Toff-C & \#H  & \#Toff-H  \\ \hline
			SIMON48/72 & 36 & 792 & 3312 & 6048 & 0 &1728 & 864 & 12744 & 6048 & 432 & 1512 & 120 \\
			SIMON48/72 & 19 & 384 & 1680 & 3192 & 0 &912 & 456 & 6624 & 3192 & 228 & 798 & 120 \\      
			SIMON48/96 & 36 & 768 & 4800 & 6048 & 0 &1728 & 864 & 14208 & 6048 & 432 & 1512 & 144 \\      
			SIMON48/96 & 19 & 360 & 2352 & 3192 & 0 &912 & 456 & 7272 & 3192 & 228 & 798 & 144 \\      
			SIMON64/96 & 42 & 1248 & 5184 & 9408 & 0 &2688 & 1344 & 19872 & 9408 & 630 & 2184 & 160 \\ 
			SIMON64/96 & 26 & 736 & 3136 & 5824 & 0 &1664 & 832 & 12192 & 5824 & 390 & 1352 & 160 \\ 
			SIMON64/128 & 44 & 1216 & 7396 & 9856 & 0 &2816 & 1408 & 22692 & 9856 & 630 & 2184 & 192 \\ 
			SIMON64/128 & 26 & 704 & 4480 & 6654 & 0 &1664 & 832 & 8184 & 3192 & 390 & 1352 & 192 \\ 
			\hline		
		\end{tabular}
	}
	\label{tb:simonimpres2}
\end{table}
\subsection{The Circuit Complexity of $\mathcal{QMKS}$ on 19-round SIMON48 and 26-round SIMON64}
\label{app2:sec2}
The circuit complexity of quantum master key exhaustive search on SIMON48 and SIMON64 given by Anand et al. is listed in Table \ref{tb:qmkrressum3}. Through the analysis process similar to Section \ref{sec:qmks}, the circuit complexity of quantum master key exhaustive search attack on these four SIMON variants can be summarized in Table \ref{tb:qmkrressum2}. After comparing the data items in Table \ref{tb:qmkrressum3} and Table \ref{tb:qmkrressum2}, we can find that in our estimate, Clifford gate count is more accurate, and T gate count, T-depth and Full-depth are lower.\\
\begin{table}[!htbp]
	\renewcommand\arraystretch{1.5}
	\centering
	\caption{\centering The circuit complexity of quantum exhaustive search on SIMON48 and SIMON64 in \cite{amm20}}
	\resizebox{\textwidth}{!}{
		\begin{tabular}{ccccccccccccc}
			\hline
			
			\multirow{2}{*}{Cipher} &\multirow{2}{*}{Round} &\multirow{2}{*}{\#NOT}  & 	\multicolumn{2}{c}{\#$CNOT_{sum}$} & \multicolumn{2}{c}{\#$H_{sum}$}&\multirow{2}{*}{\#Toff-S} &\multirow{2}{*}{\#Cliff} & \multirow{2}{*}{\#T}
			&\multirow{2}{*}{T-depth}
			&\multirow{2}{*}{Full-depth} &\multirow{2}{*}{\#qubit} \\
			
			\cmidrule(lr){4-5} 
			\cmidrule(lr){6-7} 
			& & & \#CNOT & \#Toff-C & \#H & \#Toff-H &  \\ \hline
			SIMON48/72 & 36 & $1.23\cdot2^{47}$ & $1.23 \cdot 2^{49}$ & 0 & 0 & 0 & 0 & $1.01 \cdot 2^{49.65}$ & $1.03 \cdot 2^{50.45}$ &$1.01\cdot2^{49.4}$ &$1.05\cdot2^{50.37}$ & 169\\ 
			SIMON48/96 & 36 & $1.74\cdot2^{58}$ & $1.41 \cdot 2^{62}$ & 0 & 0 & 0 & 0 & $1.02 \cdot 2^{62.66}$ & $1.02 \cdot 2^{63.05}$ &$1.01\cdot2^{61.97}$ &$1.02\cdot2^{63.11}$ & 241\\
			SIMON64/96 & 42 & $1.87\cdot2^{59}$ & $2^{62}$ & 0 & 0 & 0 & 0 & $1.02 \cdot 2^{62.27}$ & $1.02 \cdot 2^{63.08}$ &$1.01\cdot2^{61.9}$ &$1.07\cdot2^{63}$ & 224\\
			SIMON64/128 & 44 &  $1.41\cdot2^{76}$ & $1.07\cdot2^{79}$ & 0 & 0 & 0 & 0 & $1.02 \cdot 2^{79.27}$ & $1.02 \cdot 2^{79.7}$ & $1.01 \cdot 2^{78.6}$ &$1.07\cdot2^{79.8}$ & 511\\
			\hline	
		\end{tabular}
	}
	\label{tb:qmkrressum3}
\end{table}

\begin{table}[!htbp]
	\renewcommand\arraystretch{1.5}
	\centering
	\caption{The circuit complexity of quantum exhaustive search on SIMON48 and SIMON64}
	\resizebox{\textwidth}{!}{
		\begin{tabular}{ccccccccccccc}
			\hline
			
			\multirow{2}{*}{Cipher} &\multirow{2}{*}{Round} &\multirow{2}{*}{\#NOT}  & 	\multicolumn{2}{c}{\#$CNOT_{sum}$} & \multicolumn{2}{c}{\#$H_{sum}$}&\multirow{2}{*}{\#Toff-S} &\multirow{2}{*}{\#Cliff} & \multirow{2}{*}{\#T}
			&\multirow{2}{*}{T-depth}
			&\multirow{2}{*}{Full-depth} &\multirow{2}{*}{\#qubit} \\
			
			\cmidrule(lr){4-5} 
			\cmidrule(lr){6-7} 
			& & & \#CNOT & \#Toff-C & \#H & \#Toff-H &  \\ \hline
			SIMON48/72 & 36 & $1.23\cdot2^{46}$ & $1.23 \cdot 2^{49}$ & $1.23 \cdot 2^{50}$ & $1.74 \cdot 2^{42}$ & $1.41 \cdot 2^{48}$ & $1.41 \cdot 2^{47}$ & $1.23 \cdot 2^{53}$ & $1.23 \cdot 2^{50}$ &$1.41\cdot2^{46}$ &$1.23\cdot2^{48}$ & 263\\
			SIMON48/72 & 19 & $1.15\cdot2^{46}$ & $1.32 \cdot 2^{48}$ & $1.41 \cdot 2^{49}$ & $1.74 \cdot 2^{42}$ & $1.62 \cdot 2^{47}$ & $1.62 \cdot 2^{46}$ & $1.41 \cdot 2^{50}$ & $1.41 \cdot 2^{49}$ &$1.15\cdot2^{46}$ &$1.87\cdot2^{47}$ & 263\\
			SIMON48/96 & 36 & $1.15\cdot2^{58}$ & $1.41 \cdot 2^{62}$ & $1.74 \cdot 2^{62}$ & $1.15 \cdot 2^{55}$ & $2^{61}$ & $2^{59}$ & $2^{64}$ & $1.74 \cdot 2^{62}$ &$1.74\cdot2^{58}$ &$1.52\cdot2^{60}$ & 383\\ 
			SIMON48/96 & 19 & $1.15\cdot2^{58}$ & $1.32 \cdot 2^{61}$ & $1.07 \cdot 2^{62}$ & $1.15 \cdot 2^{55}$ & $1.23 \cdot 2^{60}$ & $1.23 \cdot 2^{59}$ & $1.15 \cdot 2^{63}$ & $1.07 \cdot 2^{62}$ &$1.41\cdot2^{58}$ &$1.23\cdot2^{60}$ & 383\\
			SIMON64/96 & 42 & $1.87\cdot2^{58}$ & $2^{62}$ & $2^{63}$ & $1.15 \cdot 2^{55}$ & $1.15 \cdot 2^{61}$ & $1.15 \cdot 2^{60}$ & $2^{64}$ & $2^{63}$ &$1.23\cdot2^{60}$ &$1.41\cdot2^{60}$ & 351\\
			SIMON64/96 & 26 & $1.15\cdot2^{59}$ & $2^{62}$ & $1.23 \cdot 2^{62}$ & $1.15 \cdot 2^{55}$ & $1.41 \cdot 2^{60}$ & $1.41 \cdot 2^{59}$ & $1.52 \cdot 2^{63}$ & $1.23 \cdot 2^{62}$ &$1.62\cdot2^{58}$ &$1.41\cdot2^{60}$ & 351\\
			SIMON64/128 & 44 &  $1.87\cdot2^{74}$ & $1.07\cdot2^{79}$ & $1.52 \cdot 2^{79}$ & $1.62 \cdot 2^{71}$ & $1.77 \cdot 2^{76}$ &$1.74\cdot2^{76}$ & $1.62 \cdot 2^{80}$ & $1.52 \cdot 2^{79}$ & $1.23 \cdot 2^{75}$ &$2^{77}$ & 511\\
			SIMON64/128 & 26 &  $1.62\cdot2^{75}$ & $1.32\cdot2^{78}$ & $1.87 \cdot 2^{78}$ & $1.62 \cdot 2^{71}$ & $1.07 \cdot 2^{77}$ &$1.07\cdot2^{76}$ & $1.07 \cdot 2^{80}$ & $1.87 \cdot 2^{78}$ & $2^{75}$ &$1.74\cdot2^{76}$ & 511\\
			\hline	
		\end{tabular}
	}
	\label{tb:qmkrressum2}
\end{table}

	\subsection{The Circuit Complexity of $\mathcal{QRKR}$ on 19-round SIMON48 and 26-round SIMON64}
\label{app2:sec3}
Biryukov et al. gave a 15-round differential path of SIMON48 and a 21-round differential path of SIMON64 in Table 5 of \cite{brv14} and gave the process of key recovery attacks on 19-round SIMON48 and 26-round SIMON96 respectively. Based on this result, we analyzed the quantum circuit complexity of the two stages of the quantum key recovery attack on 19-round SIMON48 and 26-round SIMON64 and listed the quantum circuit complexity in Table \ref{tb:qrkrstep3res2} and Table \ref{tb:qrkrstep4res2}.\\
\indent To carry out a quantum key recovery attack on 19-round SIMON48/72, we pick $2^{42.5}$ pairs of plaintexts. Then the process of filtering reduces the number of plaintexts into $2^{17.5}$. The first QAA instance corresponding to the quantum partial key guessing stage takes $2^{17.5}$ plaintext pairs and all 30 key bit guesses as input and outputs the superposition state of $(2^{17.5}\times2^{30})/2^{23}=2^{24.5}$ plaintext-key pair after $\lfloor \frac{\pi}{4}\sqrt{2^{23}}\rfloor$ iterations. After running this QAA instance $2^{24.5}$ times with measurements, we can expect to get $2^{24.5}[1-(1-\frac{1}{2^{24.5}})^{2^{24.5}}]\approx2^{23.8}$ different candidate keys and the correct candidate key lies in them. Similar to the circuit design shown in Fig. \ref{fig:hcircuit}, we analyzed that there are 366 CNOT gates and 124 Toffoli gates in the circuit of computing the 16th-round output difference with T-depth 36 and Full-depth 130. The second QAA instance corresponding to quantum remaining keys search stage searches the unique correct key among $2^{23.8}\times2^{42}=2^{65.8}$ keys. Similarly, we can analyze the circuit complexity of quantum key recovery attack on 19-round SIMON48/96.\\
\indent To carry out a quantum key recovery attack on 26-round SIMON64/96, we pick $2^{60.5}$ pairs of plaintexts. The process of filtering reduces the number of plaintext pairs into $2^{25.5}$ pairs. The first QAA instance corresponding to the quantum partial key guessing stage takes $2^{25.5}$ plaintexts and all 52 key bit guesses as input and outputs a superposition state of  $(2^{25.5}\times2^{52})/2^{29}=2^{48.5}$ plaintext-key pair. After running this QAA instance $2^{48.5}$ times with measurements, we expect to get $2^{48.5}[1-(1-\frac{1}{2^{48.5}})^{2^{48.5}}]\approx2^{47.8}$ different candidate keys, and the correct candidate key lies in it. Similar to the circuit design shown in Fig. \ref{fig:hcircuit}, we analyzed that there are 560 CNOT gates and 207 Toffoli gates in the circuit of obtaining 22nd-round output differential with T-depth 51 and Full-depth 185. The second QAA instance corresponding to quantum remaining keys search stage searchs the unique and correct key among $2^{47.8}\times2^{44}=2^{91.8}$ keys. Similarly, we can analyze the circuit complexity of the quantum key recovery attack on 26-round SIMON64/128.
\begin{table}[!htbp]
	\renewcommand\arraystretch{1.5}
	\centering
	\caption{The circuit complexity of quantum partial key guessing on SIMON48 and SIMON64}
	\resizebox{\textwidth}{!}{
		\begin{tabular}{ccccccccccccc}
			\hline
			\multirow{2}{*}{Cipher} & \multirow{2}{*}{Round} &\multirow{2}{*}{\#NOT}  & \multicolumn{2}{c}{\#$CNOT_{sum}$} & \multicolumn{2}{c}{\#$H_{sum}$}&\multirow{2}{*}{\#Toff-S} &\multirow{2}{*}{\#Cliff} & \multirow{2}{*}{\#T} &\multirow{2}{*}{T-depth}&\multirow{2}{*}{Full-depth}&\multirow{2}{*}{\#qubit} \\
			
			\cmidrule(lr){4-5} 
			\cmidrule(lr){6-7} 
			& & & \#CNOT & \#Toff-C & \#H & \#Toff-H &  \\ \hline
			SIMON48/72 & 19 & 0 & $1.62 \cdot 2^{20}$ & $1.87 \cdot 2^{22}$ & $1.07 \cdot 2^{17}$ & $1.07 \cdot 2^{21}$ & $1.07 \cdot 2^{20}$ & $1.52 \cdot 2^{23}$ & $1.87 \cdot 2^{22}$ & $1.23 \cdot 2^{45}$ & $1.07 \cdot 2^{47}$ & 347\\    
			SIMON48/96 & 19 & 0 & $1.62 \cdot 2^{20}$ & $1.87 \cdot 2^{22}$ & $1.07 \cdot 2^{17}$ & $1.07 \cdot 2^{21}$ & $1.07 \cdot 2^{20}$ & $1.52 \cdot 2^{23}$ & $1.87 \cdot 2^{22}$ & $1.23 \cdot 2^{45}$ & $1.07 \cdot 2^{47}$ & 347\\ 
			SIMON64/96 & 26 & 0 & $1.23 \cdot 2^{24}$ & $1.41 \cdot 2^{26}$ & $1.87 \cdot 2^{20}$ & $1.62 \cdot 2^{24}$ & $1.62 \cdot 2^{23}$ & $1.23 \cdot 2^{27}$ & $1.41 \cdot 2^{26}$ & $1.74 \cdot 2^{73}$ & $1.62 \cdot 2^{74}$ & 487\\
			SIMON64/128 & 26 & 0 & $1.23 \cdot 2^{24}$ & $1.41 \cdot 2^{26}$ & $1.87 \cdot 2^{20}$ & $1.62 \cdot 2^{24}$ & $1.62 \cdot 2^{23}$ & $1.23 \cdot 2^{27}$ & $1.41 \cdot 2^{26}$ & $1.74 \cdot 2^{73}$ & $1.62 \cdot 2^{74}$ & 487\\
			
			\hline	
		\end{tabular}
	}
	\label{tb:qrkrstep3res2}
\end{table}

\begin{table}[!htbp]
	\renewcommand\arraystretch{1.5}
	\centering
	\caption{The circuit complexity of quantum remaining keys search on SIMON48 and SIMON64}
	\resizebox{\textwidth}{!}{
		\begin{tabular}{ccccccccccccc}
			\hline
			\multirow{2}{*}{Cipher} &\multirow{2}{*}{Round} &\multirow{2}{*}{\#NOT}  & \multicolumn{2}{c}{\#$CNOT_{sum}$} & 	\multicolumn{2}{c}{\#$H_{sum}$} &\multirow{2}{*}{\#Toff-S} &\multirow{2}{*}{\#Cliff} & \multirow{2}{*}{\#T}
			& \multirow{2}{*}{T-depth}& \multirow{2}{*}{Full-depth} &\multirow{2}{*}{\#qubit} \\ 
			
			\cmidrule(lr){4-5} 
			\cmidrule(lr){6-7} 
			& & & \#CNOT & \#Toff-C & \#H & \#Toff-H &  \\ \hline
			SIMON48/72 & 19 & $1.07 \cdot 2^{43}$ & $1.23 \cdot 2^{45}$ & $1.32 \cdot 2^{46}$ & $1.87 \cdot 2^{38}$ & $1.52 \cdot 2^{44}$ & $1.52 \cdot 2^{43}$ & $1.32 \cdot 2^{47}$ & $1.32 \cdot 2^{46}$ & $1.07 \cdot 2^{43}$ & $1.74 \cdot 2^{44}$ & 263\\
			SIMON48/96 & 19 & $2^{55}$ & $1.74 \cdot 2^{57}$  & $1.41 \cdot 2^{58}$ & $1.52 \cdot 2^{51}$ & $1.62 \cdot 2^{55}$ & $1.87 \cdot 2^{56}$ &$1.52 \cdot 2^{59}$ & $1.41 \cdot 2^{58}$& $1.15 \cdot 2^{55}$ &$2^{57}$ & 287\\
			SIMON64/96 & 26 & $1.07 \cdot 2^{57}$ & $1.15 \cdot 2^{59}$ & $1.32 \cdot 2^{59}$ & $2^{52}$ & $1.32 \cdot 2^{58}$ & $1.32 \cdot 2^{57}$ & $1.87 \cdot 2^{60}$ & $1.23 \cdot 2^{61}$ & $1.52 \cdot 2^{56}$ & $1.32 \cdot 2^{58}$ & 351\\
			SIMON64/128 & 26 & $1.23 \cdot 2^{73}$ & $1.62 \cdot 2^{75}$ & $1.23 \cdot 2^{76}$ & $1.74\cdot2^{68}$ & $1.41 \cdot 2^{74}$ & $1.41 \cdot 2^{73}$ & $1.32 \cdot 2^{77}$ & $1.23 \cdot 2^{76}$ & $1.62 \cdot 2^{72}$ & $1.41 \cdot 2^{74}$ & 368\\
			
			\hline
			
		\end{tabular}
	}
	\label{tb:qrkrstep4res2}
\end{table}

\subsection{The Complexity Summary of $\mathcal{QMKS}$ and $\mathcal{QRKR}$ on SIMON Block Cipher }
To summarize more clearly, we compare the encryption complexity and circuit complexity of the quantum master key exhaustive search attack on 19-round SIMON32, 19-round SIMON48 and 26-round SIMON64 and quantum round key recovery attack on these SIMON variants in Table \ref{tb:comp}. \\
\indent We can observe that the quantum round key recovery attack on 26-round SIMON64/96 doesn't work, because the encryption complexity and circuit depth of this attack are much higher than those of the quantum master key search attack. By comparing the third row of Table \ref{tb:qrkrstep3res2} with the third row of Table \ref{tb:qrkrstep4res2}, it can be seen that the reason for the failure of the attack is that the first QAA instance needs to be run many times to generate candidate keys. As a result, the circuit depth of the first stage is much greater than that of the second stage. For other SIMON variants, the encryption complexity and circuit complexity of the quantum round key recovery attack are lower than those of the quantum master key search attack.

\begin{table}[h]
		\renewcommand\arraystretch{1.5}
	
	\caption{The Comparison of the complexity of quantum master key search attack and quantum round key recovery attack}
	\resizebox{\textwidth}{!}{
		\begin{threeparttable}  
			\begin{tabular}{ccccccccc}
				\hline
				\multirow{2}{*}{Cipher} &\multirow{2}{*}{Round} &\multirow{2}{*}{Technique} &\multirow{2}{*}{E. C.}  & 	\multicolumn{5}{c}{C. C.} \\
				\cmidrule(lr){5-9} 
				& & & & \#Cliff & \#T & T-depth  & Full-depth & \#qubit  \\ \hline
				
				SIMON32/64 & 19 & $\mathcal{QMKS}$ &$2^{32.6}$ & $1.32\cdot 2^{46}$ & $1.41\cdot 2^{45}$ & $2^{42}$ & $1.74\cdot2^{43}$ & 255 \\
				SIMON32/64 & 19 & $\mathcal{QRKR}$ &$2^{31.1}$ & $1.15\cdot 2^{41}$ & $1.52\cdot 2^{40}$ & $1.52\cdot2^{39}$ & $1.32\cdot2^{41}$ & 400\\
				SIMON48/72 & 19 & $\mathcal{QMKS}$ & $2^{36.6}$ & $1.41\cdot2^{50}$ & $1.41\cdot2^{49}$ & $1.15\cdot2^{46}$ & $1.87\cdot2^{47}$ &  263\\      
				SIMON48/72 & 19 & $\mathcal{QRKR}$ & $2^{34.8}$ & $1.32\cdot2^{47}$ & $1.32\cdot2^{46}$ & $1.52\cdot2^{45}$ & $1.32\cdot2^{47}$ &  610\\
				SIMON48/96 & 19 & $\mathcal{QMKS}$ & $2^{48.6}$ & $1.15\cdot2^{63}$ & $1.07\cdot2^{62}$ & $1.41\cdot2^{58}$ & $1.23\cdot2^{60}$ &  383\\
				SIMON48/96 & 19 & $\mathcal{QRKR}$ & $2^{45.6}$ & $1.52\cdot2^{59}$ & $1.41\cdot2^{58}$ & $1.15\cdot2^{55}$ & $2^{57}$ &  634\\
				SIMON64/96 & 26 & $\mathcal{QMKS}$ & $2^{48.6}$ & $1.52\cdot2^{63}$ & $1.23\cdot2^{62}$ & $1.62\cdot2^{58}$ & $1.41\cdot2^{60}$ &  351 \\ 
				SIMON64/96 & 26 & $\mathcal{QRKR}$ & $2^{60.5}$ & $1.87\cdot2^{60}$ & $1.23\cdot2^{61}$ & $1.74\cdot2^{73}$ & $1.62\cdot2^{74}$ &  838 \\
				SIMON64/128 & 26 & $\mathcal{QMKS}$ & $2^{64.6}$ & $1.07\cdot2^{80}$ & $1.87\cdot2^{78}$ & $2^{75}$ & $1.74\cdot2^{76}$ &  511 \\
				SIMON64/128 & 26 & $\mathcal{QRKR}$ & $2^{62.8}$ & $1.32\cdot2^{77}$ & $1.23\cdot2^{76}$ & $1.23 \cdot 2^{74}$ & $1.52\cdot2^{75}$ &  855 \\ 
				\hline		
			\end{tabular}
			 \begin{tablenotes}    
				\footnotesize               
				\item[1] E.C. is the abbreviation of Encryption Complexity.
				\item[2] C.C. is the abbreviation of Circuit Complexity.
			\end{tablenotes}            
	\end{threeparttable}       
	}
	\label{tb:comp}
\end{table}

	\end{subappendices}

	\bibliographystyle{abbrv}
	\bibliography{main}

\end{document}